\documentclass{article}
\usepackage{amssymb}
\usepackage{amsmath}
\usepackage{amsfonts}
\usepackage{sw20ssur}
\usepackage{cite}
\usepackage{afterpage}

\newtheorem{theorem}{Theorem}
\newtheorem{acknowledgement}[theorem]{Acknowledgement}

\input{tcilatex}
\begin{document}

\title{Probabilistic analysis of three-player symmetric quantum games played
using the Einstein-Podolsky-Rosen-Bohm setting}
\author{Azhar Iqbal$^{\text{a,b}}$, Taksu Cheon$^{\text{c}}$, and Derek
Abbott$^{\text{a }}$ \\
$^{\text{a}}${\small School of Electrical \& Electronic Engineering, The
University of Adelaide, SA 5005, Australia.}\\
$^{\text{b}}${\small Centre for Advanced Mathematics and Physics, National
University of Sciences \& Technology,}\\
{\small Campus of College of Electrical \& Mechanical Engineering, Peshawar
Road, Rawalpindi, Pakistan.}\\
$^{\text{c}}${\small Kochi University of Technology, Tosa Yamada, Kochi
782-8502, Japan.}}
\maketitle
\tableofcontents

\begin{abstract}
This paper extends our probabilistic framework for two-player quantum games
to the mutliplayer case, while giving a unified perspective for both
classical and quantum games. Considering joint probabilities in the standard
Einstein-Podolsky-Rosen-Bohm (EPR-Bohm) setting for three observers, we use
this setting in order to play general three-player non-cooperative symmetric
games. We analyze how the peculiar non-factorizable joint probabilities
provided by the EPR-Bohm setting can change the outcome of a game, while
requiring that the quantum game attains a classical interpretation for
factorizable joint probabilities. In this framework, our analysis of the
three-player generalized Prisoner's Dilemma (PD) shows that the players can
indeed escape from the classical outcome of the game, because of
non-factorizable joint probabilities that the EPR setting can provide. This
surprising result for three-player PD contrasts strikingly with our earlier
result for two-player PD, played in the same framework, in which even
non-factorizable joint probabilities do not result in escaping from the
classical consequence of the game.
\end{abstract}

\newpage

\section{\label{Intro}Introduction}

Although multiplayer games have been extensively studied in the game theory
literature \cite{Rasmusen,Khan} their analysis is often found more complex
than for two-player games \cite{Constantinos}. Economics \cite{Aumann} and
mathematical biology \cite{Broom(a),Broom(b),HofbauerSigmund} are the areas
where most applications of multiplayer games are discussed.

Quantum games first came into prominence following the work of Meyer \cite%
{MeyerDavid} and Eisert et al \cite{EWL}. However, the first game-like
situations involving many agents were brought to the quantum regime in 1990
by Mermin \cite{Mermin,Mermin1}. Mermin analyzed a multiplayer game that can
be won with certainty when it is played using spin-half particles in a
Greenberger-Horne-Zeilinger (GHZ) state \cite{GHZ}, while no classical
strategy can achieve this.

In 1999, Vaidman \cite{Vaidman,Vaidman1} described the GHZ paradox \cite{GHZ}
using a game among three players. Vaidman's game is now well known in the
quantum game literature. In a similar vein, others \cite%
{AharonovRohrlich,Fulton} have discussed game-like situations involving many
players for which quantum mechanics significantly helps their chances of
winning.

The motivation behind these developments \cite{AbbottShaliziDavies} was to
use a game framework in order to demonstrate the remarkable, and often
counterintuitive, quantum correlations that may arise when many agents
interact while sharing quantum resources.

Systematic procedures were suggested \cite%
{BenjaminHayden,MarinattoWeber,Johnson,Piotrowski,Du,Flitney,FlitneyAbbottA,FlitneyAbbottB,FlitneyAbbottC,Cheon,Shimamura,IqbalCheonA,IqbalToor1,IqbalToor2,IqbalToor3,IqbalWeigert}
to quantize a given game and earlier work considered noncooperative games in
their normal-form \cite{Rasmusen}.

The approach towards playing quantum games is distinct in that, instead of
inventing games in which quantum correlations help players winning games, it
proposes quantization procedures for given and very often well known games.
That is, instead of tailoring `winning conditions' for the invented games,
which can be satisfied when the game is played by quantum players, quantum
game theory finds how the sharing of quantum resources may
replace/displace/change the solution(s) or the outcome(s) of known games.
The emphasis is, therefore, shifted from inventing games to the writing of
prescription(s) for quantizing well-known games.

In the area of quantum games, multiplayer games were first studied by
Benjamin and Hayden \cite{BenjaminHayden} and have subsequently been
considered by many others \cite%
{KayJohnson,DuLi,HanZhang,IqbalToor,FlitneyAbbott,ChenWang,FlitneyHollenberg,FlitneyHollenberg1}%
.

This paper extends to multiplayer quantum games the probabilistic framework
originally proposed for two-player noncooperative games by present authors
in Ref.~\cite{IqbalCheon}. This framework unifies classical and quantum
two-player games while using the Einstein-Podolsky-Rosen-Bohm (EPR-Bohm)
experiments \cite{EPR,Bell,Bell(a),Bell1,Bell2,Aspect,Peres,Cereceda} to
play a two-player game.

It has been reported in literature \cite{Fine,WinsbergFine,Fine1} that joint
probabilities in EPR-Bohm experiments may not be factorizable. This led us
in Refs.~\cite{IqbalCheon,IqbalCheonM} to construct quantum games from
non-factorizable joint probabilities. To ensure that the classical game
remains embedded within the corresponding quantum game, this framework
requires that quantum game attains classical interpretation when joint
probabilities are factorizable.

As this framework proposes an entirely probabilistic argument in the
construction of quantum games it thus provides a unifying perspective for
both classical and quantum games. It also presents a more accessible
analysis of quantum games, which can be of potential interest to readers
outside the quantum physics domain.

\section{\label{SG}Three-player symmetric games}

We consider three-player symmetric (noncooperative) games in which three
players (henceforth labelled as Alice, Bob, and Chris) make their choices
simultaneously. The players are assumed located at distance and that they
are unable to communicate among themselves. They, however, can communicate
to a referee who organizes the game and ensures that the rules of the game
are obeyed by the participating parties.

We assume that each player has two choices that we refer to as his/her pure
strategies. The payoff relations are made public by the referee at the start
of the game. At some particular instant the players are asked to inform the
referee of their strategies. The players' payoffs depend on the game they
play, their strategies, and the set of joint probabilities relevant to the
probabilistic system which they share in order to play the game.

We assume Alice's strategies are $S_{1},$ $S_{2}$; Bob's strategies are $%
S_{1}^{\prime },$ $S_{2}^{\prime }$; and Chris' strategies are $%
S_{1}^{\prime \prime },$ $S_{2}^{\prime \prime },$ and that we have the
following pure-strategy payoff relations \cite{IqbalToor} that define the
game

\begin{equation}
\begin{array}{l}
\Pi _{A,B,C}(S_{1},S_{1}^{\prime },S_{1}^{\prime \prime })=\alpha _{1},\beta
_{1},\gamma _{1}; \\ 
\Pi _{A,B,C}(S_{2},S_{1}^{\prime },S_{1}^{\prime \prime })=\alpha _{2},\beta
_{2},\gamma _{2}; \\ 
\Pi _{A,B,C}(S_{1},S_{2}^{\prime },S_{1}^{\prime \prime })=\alpha _{3},\beta
_{3},\gamma _{3}; \\ 
\Pi _{A,B,C}(S_{1},S_{1}^{\prime },S_{2}^{\prime \prime })=\alpha _{4},\beta
_{4},\gamma _{4};%
\end{array}%
\begin{array}{l}
\Pi _{A,B,C}(S_{1},S_{2}^{\prime },S_{2}^{\prime \prime })=\alpha _{5},\beta
_{5},\gamma _{5}; \\ 
\Pi _{A,B,C}(S_{2},S_{1}^{\prime },S_{2}^{\prime \prime })=\alpha _{6},\beta
_{6},\gamma _{6}; \\ 
\Pi _{A,B,C}(S_{2},S_{2}^{\prime },S_{1}^{\prime \prime })=\alpha _{7},\beta
_{7},\gamma _{7}; \\ 
\Pi _{A,B,C}(S_{2},S_{2}^{\prime },S_{2}^{\prime \prime })=\alpha _{8},\beta
_{8},\gamma _{8},%
\end{array}
\label{payoffs constants definitions}
\end{equation}%
where the subscripts $A$, $B$, and $C$ refer to Alice, Bob, and Chris,
respectively.

In one of these payoff relations, the three entries in brackets on left side
are Alice's, Bob's, and Chris' pure strategies, respectively, and the three
numbers on the right side are their rewards, respectively. For example, $\Pi
_{A,B,C}(S_{1},S_{2}^{\prime },S_{1}^{\prime \prime })=\alpha _{3},$ $\beta
_{3},$ $\gamma _{3}$ states that Alice, Bob, and Chris obtain $\alpha _{3},$ 
$\beta _{3},$ $\gamma _{3}$, respectively, when they play the strategies $%
S_{1},$ $S_{2}^{\prime },$ $S_{1}^{\prime \prime }$, respectively.

In repeated runs, a player can choose between his/her two pure strategies
with some probability, which defines his/her mixed-strategy. We denote a
mixed-strategy by $x,$ $y,$ $z\in \lbrack 0,1]$ for Alice, Bob, and Chris,
respectively. These are probabilities with which Alice, Bob, and Chris play
the pure strategies $S_{1},$ $S_{1}^{\prime },$ $S_{1}^{\prime \prime }$,
respectively. They, then, play the pure strategies $S_{2},$ $S_{2}^{\prime
}, $ $S_{2}^{\prime \prime }$ with probabilities $(1-x),$ $(1-y)$, $(1-z)$,
respectively, and the mixed-strategy payoff relations, therefore, read

\begin{equation}
\begin{array}{l}
\Pi _{A,B,C}(x,y,z)=xyz\Pi _{A,B,C}(S_{1},S_{1}^{\prime },S_{1}^{\prime
\prime })+x(1-y)z\Pi _{A,B,C}(S_{1},S_{2}^{\prime },S_{1}^{\prime \prime })+
\\ 
xy(1-z)\Pi _{A,B,C}(S_{1},S_{1}^{\prime },S_{2}^{\prime \prime
})+x(1-y)(1-z)\Pi _{A,B,C}(S_{1},S_{2}^{\prime },S_{2}^{\prime \prime })+ \\ 
(1-x)yz\Pi _{A,B,C}(S_{2},S_{1}^{\prime },S_{1}^{\prime \prime
})+(1-x)(1-y)z\Pi _{A,B,C}(S_{2},S_{2}^{\prime },S_{1}^{\prime \prime })+ \\ 
(1-x)y(1-z)\Pi _{A,B,C}(S_{2},S_{1}^{\prime },S_{2}^{\prime \prime
})+(1-x)(1-y)(1-z)\Pi _{A,B,C}(S_{2},S_{2}^{\prime },S_{2}^{\prime \prime }).%
\end{array}
\label{3-coin mixed-strategy payoffs}
\end{equation}

In this paper we consider symmetric three-player games that are defined by
the condition that a player's payoff is decided by his/her strategy and not
by his/her identity. Mathematically, this is expressed by the conditions

\begin{equation}
\Pi _{A}(x,y,z)=\Pi _{A}(x,z,y)=\Pi _{B}(y,x,z)=\Pi _{B}(z,x,y)=\Pi
_{C}(y,z,x)=\Pi _{C}(z,y,x).  \label{constraints for symmetric game}
\end{equation}%
The payoff relations (\ref{3-coin mixed-strategy payoffs}) satisfy these
conditions (\ref{constraints for symmetric game}) when \cite{IqbalToor},

\begin{equation}
\begin{array}{c}
\begin{array}{cccc}
\beta _{1}=\alpha _{1}, & \beta _{2}=\alpha _{3}, & \beta _{3}=\alpha _{2},
& \beta _{4}=\alpha _{3}, \\ 
\beta _{5}=\alpha _{6}, & \beta _{6}=\alpha _{5}, & \beta _{7}=\alpha _{6},
& \beta _{8}=\alpha _{8}, \\ 
\gamma _{1}=\alpha _{1}, & \gamma _{2}=\alpha _{3}, & \gamma _{3}=\alpha
_{3}, & \gamma _{4}=\alpha _{2}, \\ 
\gamma _{5}=\alpha _{6}, & \gamma _{6}=\alpha _{6}, & \gamma _{7}=\alpha
_{5}, & \gamma _{8}=\alpha _{8},%
\end{array}
\\ 
\begin{array}{cc}
\alpha _{6}=\alpha _{7}, & \alpha _{3}=\alpha _{4}.%
\end{array}%
\end{array}%
\end{equation}%
A symmetric three-player game can, therefore, be defined by only six
constants $\alpha _{1},$ $\alpha _{2},$ $\alpha _{3},$ $\alpha _{5},$ $%
\alpha _{6},$ and $\alpha _{8}$. In the rest of this paper we will define
these six constants to be $\alpha ,$ $\beta ,$ $\delta ,$ $\epsilon ,$ $%
\theta ,$ $\omega $ where $\alpha _{1}=\alpha ,$ $\alpha _{2}=\beta ,$ $%
\alpha _{3}=\delta ,$ $\alpha _{5}=\epsilon ,$ $\alpha _{6}=\theta ,$ and $%
\alpha _{8}=\omega $. The pure-strategy payoff relations (\ref{payoffs
constants definitions}) are then re-expressed as

\begin{equation}
\begin{array}{l}
\Pi _{A,B,C}(S_{1},S_{1}^{\prime },S_{1}^{\prime \prime })=\alpha ,\alpha
,\alpha ; \\ 
\Pi _{A,B,C}(S_{2},S_{1}^{\prime },S_{1}^{\prime \prime })=\beta ,\delta
,\delta ; \\ 
\Pi _{A,B,C}(S_{1},S_{2}^{\prime },S_{1}^{\prime \prime })=\delta ,\beta
,\delta ; \\ 
\Pi _{A,B,C}(S_{1},S_{1}^{\prime },S_{2}^{\prime \prime })=\delta ,\delta
,\beta ;%
\end{array}%
\begin{array}{l}
\Pi _{A,B,C}(S_{1},S_{2}^{\prime },S_{2}^{\prime \prime })=\epsilon ,\theta
,\theta ; \\ 
\Pi _{A,B,C}(S_{2},S_{1}^{\prime },S_{2}^{\prime \prime })=\theta ,\epsilon
,\theta ; \\ 
\Pi _{A,B,C}(S_{2},S_{2}^{\prime },S_{1}^{\prime \prime })=\theta ,\theta
,\epsilon ; \\ 
\Pi _{A,B,C}(S_{2},S_{2}^{\prime },S_{2}^{\prime \prime })=\omega ,\omega
,\omega .%
\end{array}
\label{symmetric 3-player game definition}
\end{equation}

\subsection{Three-player Prisoner's Dilemma}

Prisoner's Dilemma (PD) is a noncooperative game \cite{Rasmusen} that is
widely known in the areas of economics, social, and political sciences. In
recent years quantum physics has been added to this list. This game was
investigated \cite{EWL} early in the history of quantum games and is known
to have had provided significant motivation for further work in this area.

Two-player PD is about two criminals (referred hereafter as players) who are
arrested after having committed a crime. The investigators have the
following plan to make them confess their crime. Both are placed in separate
cells and are not allowed to communicate. They are contacted individually
and are asked to choose between two choices (strategies): \emph{to confess} $%
(D)$ and \emph{not to confess} $(C)$, where $C$ and $D$ stand for
Cooperation and Defection and this well-known wording of their available
choices refers to the fellow prisoner and not to the authorities.

The rules state that if neither prisoner confesses, i.e. $(C,C)$, both are
given freedom; when one prisoner confesses $(D)$ and the other does not $(C)$%
, i.e. $(C,D)$ or $(D,C)$, the prisoner who confesses gets freedom as well
as financial reward, while the prisoner who did not confess ends up in the
prison for a longer term. When both prisoners confess, i.e. $(D,D)$, both
are given a reduced term.

In the two-player case the strategy pair $(D,D)$ comes out as the unique NE
(and the rational outcome) of the game, leading to the situation of both
having reduced term. The game offers a dilemma as the rational outcome $%
(D,D) $ differs from the outcome $(C,C)$, which is an available choice, and
for which both prisoners get freedom.

The three-player PD is defined by making the association:

\begin{equation}
S_{1}\sim C,\text{ }S_{2}\sim D,\text{ }S_{1}^{\prime }\sim C,\text{ }%
S_{2}^{\prime }\sim D,\text{ }S_{1}^{\prime \prime }\sim C,\text{ }%
S_{2}^{\prime \prime }\sim D
\end{equation}%
and afterwards imposing the following conditions \cite{3playerPD}:

a) The strategy $S_{2}$ is a dominant choice \cite{Rasmusen} for each
player. For Alice this requires:

\begin{equation}
\begin{array}{l}
\Pi _{A}(S_{2},S_{1}^{\prime },S_{1}^{\prime \prime })>\Pi
_{A}(S_{1},S_{1}^{\prime },S_{1}^{\prime \prime }), \\ 
\Pi _{A}(S_{2},S_{2}^{\prime },S_{2}^{\prime \prime })>\Pi
_{A}(S_{1},S_{2}^{\prime },S_{2}^{\prime \prime }), \\ 
\Pi _{A}(S_{2},S_{1}^{\prime },S_{2}^{\prime \prime })>\Pi
_{A}(S_{1},S_{1}^{\prime },S_{2}^{\prime \prime }),%
\end{array}%
\end{equation}%
and similar inequalities hold for players Bob and Chris.

b) A player is better off if more of his opponents choose to cooperate. For
Alice this requires:

\begin{equation}
\begin{array}{l}
\Pi _{A}(S_{2},S_{1}^{\prime },S_{1}^{\prime \prime })>\Pi
_{A}(S_{2},S_{1}^{\prime },S_{2}^{\prime \prime })>\Pi
_{A}(S_{2},S_{2}^{\prime },S_{2}^{\prime \prime }), \\ 
\Pi _{A}(S_{1},S_{1}^{\prime },S_{1}^{\prime \prime })>\Pi
_{A}(S_{1},S_{1}^{\prime },S_{2}^{\prime \prime })>\Pi
_{A}(S_{1},S_{2}^{\prime },S_{2}^{\prime \prime }),%
\end{array}%
\end{equation}%
and similar relations hold for Bob and Chris.

c) If one player's choice is fixed, the other two players are left in the
situation of a two-player PD. For Alice this requires:

\begin{equation}
\begin{array}{l}
\Pi _{A}(S_{1},S_{1}^{\prime },S_{2}^{\prime \prime })>\Pi
_{A}(S_{2},S_{2}^{\prime },S_{2}^{\prime \prime }), \\ 
\Pi _{A}(S_{1},S_{1}^{\prime },S_{1}^{\prime \prime })>\Pi
_{A}(S_{2},S_{1}^{\prime },S_{2}^{\prime \prime }), \\ 
\Pi _{A}(S_{1},S_{1}^{\prime },S_{2}^{\prime \prime })>(1/2)\left\{ \Pi
_{A}(S_{1},S_{2}^{\prime },S_{2}^{\prime \prime })+\Pi
_{A}(S_{2},S_{1}^{\prime },S_{2}^{\prime \prime })\right\} , \\ 
\Pi _{A}(S_{1},S_{1}^{\prime },S_{1}^{\prime \prime })>(1/2)\left\{ \Pi
_{A}(S_{1},S_{1}^{\prime },S_{2}^{\prime \prime })+\Pi
_{A}(S_{2},S_{1}^{\prime },S_{1}^{\prime \prime })\right\} ,%
\end{array}%
\end{equation}%
and similar relations hold for Bob and Chris.

Translating the above conditions while using the notation introduced in (\ref%
{symmetric 3-player game definition}) requires:

\begin{equation}
\begin{array}{l}
\text{a) }\beta >\alpha ,\ \ \omega >\epsilon ,\ \ \theta >\delta \\ 
\text{b) }\beta >\theta >\omega ,\ \ \alpha >\delta >\epsilon \\ 
\text{c) }\delta >\omega ,\ \ \alpha >\theta ,\ \ \delta >(1/2)(\epsilon
+\theta ),\ \ \alpha >(1/2)(\delta +\beta )%
\end{array}
\label{Generalized PD Definition}
\end{equation}%
which define a generalized three-player PD. For example \cite{3playerPD}, by
letting $\alpha =7,$ $\beta =9,$ $\delta =3,$ $\epsilon =0,$\ $\omega =1,$ $%
\theta =5$ all of these conditions hold.

\section{Playing three-player games using coins}

We consider the situation when three players share a probabilistic system to
play the three-player symmetric game defined in the Section (\ref{SG}). For
this system, in a run, a player has to choose between one out of two pure
strategies and, in either case, the outcome (of some measurement, or
observation, which follows after players have made their choices) is either $%
+1$ or $-1$.

When we associate $+1$ with the Head and $-1$ with the Tail of a coin,
sharing coins (not necessarily unbiased) provides a physical realization of
a probabilistic physical system. In the following we consider two setups,
both of which use coins in order to play the symmetric three-player game (%
\ref{symmetric 3-player game definition}). We find that the later setup
provides an appropriate arrangement for introducing joint probabilities
(associated with an EPR-Bohm setting involving three observers) in the
playing of a three-player game.

We note that in the standard EPR-Bohm setting, in a run, each one of the
spatially-separated observers chooses one between two directions. A quantum
measurement along the two chosen directions, in a run, generate either $+1$
or $-1$ as the outcome. That is, in a run one of the four possible outcomes $%
(+1,+1,+1),$ $(+1,+1,-1),$ $(+1,-1,+1),$ $(+1,-1,-1),$ $(-1,+1,+1),$ $%
(-1,+1,-1),$ $(-1,-1,+1),$ $(-1,-1,-1)$ emerges.

\subsection{Three-coin setup}

The most natural scenario for playing a three-player game, when they share a
probabilistic physical system that involves three coins, is the one when in
a run each player is given a coin in a Head state, and `to flip' or to `not
to flip' are the player's available strategies. We denote Alice's, Bob's,
and Chris' strategy `to flip' by $S_{1},$ $S_{1}^{\prime },$ and $%
S_{1}^{^{\prime \prime }}$, respectively, and likewise, we denote Alice's,
Bob's, and Chris' strategy `not to flip' by $S_{2},$ $S_{2}^{\prime },$ and $%
S_{2}^{^{\prime \prime }},$ respectively. The three coins are then passed to
a referee who rewards players after observing the state of the three coins.

In repeated runs, the players Alice, Bob, and Chris can play mixed
strategies denoted by $x,y,z$ $\in \lbrack 0,1]$, respectively. Here $x,y,z$
are the probabilities to choose $S_{1}$ (out of $S_{1}$ and $S_{2}$), $%
S_{1}^{\prime }$ (out of $S_{1}^{\prime }$ and $S_{2}^{\prime }$), and $%
S_{1}^{\prime \prime }$ (out of $S_{1}^{\prime \prime }$ and $S_{2}^{\prime
\prime }$), by Alice, Bob, and Chris, respectively:

\begin{equation}
\begin{array}{l}
\Pi _{A,B,C}(x,y,z)=xyz(\alpha ,\alpha ,\alpha )+x(1-y)z(\delta ,\beta
,\delta )+xy(1-z)(\delta ,\delta ,\beta )+ \\ 
x(1-y)(1-z)(\epsilon ,\theta ,\theta )+(1-x)yz(\beta ,\delta ,\delta
)+(1-x)(1-y)z(\theta ,\theta ,\epsilon )+ \\ 
(1-x)y(1-z)(\theta ,\epsilon ,\theta )+(1-x)(1-y)(1-z)(\omega ,\omega
,\omega ).%
\end{array}%
\end{equation}%
Assume $(x^{\star },y^{\star },z^{\star })$ is a Nash Equilibrium (NE) \cite%
{Rasmusen} then:

\begin{equation}
\begin{array}{l}
\Pi _{A}(x^{\star },y^{\star },z^{\star })-\Pi _{A}(x,y^{\star },z^{\star
})\geqslant 0, \\ 
\Pi _{B}(x^{\star },y^{\star },z^{\star })-\Pi _{B}(x^{\star },y,z^{\star
})\geqslant 0, \\ 
\Pi _{C}(x^{\star },y^{\star },z^{\star })-\Pi _{B}(x^{\star },y^{\star
},z)\geqslant 0.%
\end{array}
\label{3-coin NE}
\end{equation}%
In the following, we will use NE\ when we refer to either a Nash Equilibrium
or to Nash Equilibria, as determined by the context. We call this
arrangement, which uses three coins to play a three-player game, the \emph{%
three-coin setup}.

\subsection{Six-coin setup}

The three-player game (\ref{symmetric 3-player game definition}) can also be
played using six coins (not necessarily unbiased) instead of the three. This
can be arranged as follows. Two coins are assigned to each player before the
game begins. In a run each player chooses one out of the two, which defines
his/her strategy in the run. Three coins are, therefore, chosen in a run.
The three chosen coins are passed to a referee who tosses them together and
observes the outcome. Many such outcomes are observed as the process of
receiving, choosing, and subsequently tossing the coins is repeated many
times.

After many runs, the referee rewards the players according to their
strategies (i.e.~which coin(s) they have chosen over many runs), the
outcomes of several tosses giving rise to the underlying statistics of the
coins and from the six coefficients defining the three-player symmetric game
defined in Section (\ref{SG}).

Notice that coins are tossed in each run, which gives the playing of a game
an inherently probabilistic character. This paves the way to step into the
quantum regime and provides the key for introducing quantum probabilities.

We call this arrangement of using six coins, for playing a three-player
game, the \emph{six-coin setup}. Why introduce a six-coin setup when a
three-player game can also be played in three-coin setup? The answer is
provided by the EPR-Bohm setting that involves three observers and $64$
joint probabilities. The six coin setup allows us to translate the playing
of a three-player game in terms of $64$ joint probabilities. When these
joint probabilities are quantum mechanical (and are obtained from an
EPR-Bohm setting involving three observers) they might have the unusual
character of being non-factorizable. That is, the six-coin setup serves as
an intermediate step allowing us to see the impact of non-factorizable
quantum probabilities on the solution of a game.

In the six-coin setup, by our definition, a player plays a pure strategy%
\footnote{%
This definition of a pure strategy, of course, corresponds to the usual
definition of a mixed-strategy, in accordance with the known result in
quantum games that a product pure state results in a mixed-strategy
classical game.} when s/he chooses the same coin over all the runs and s/he
plays a mixed strategy when s/he chooses his/her first coin with some
probability over the runs.

Notice that, by its construction, this setup requires a large number of runs
for playing a game, irrespective of whether players play the pure strategies
or the mixed strategies, as in either case players' payoffs depend on
outcomes of many tosses.

We denote Alice's two coins by $S_{1},$ $S_{2}$; Bob's two coins by $%
S_{1}^{\prime },$ $S_{2}^{\prime }$; and Chris' two coins by $S_{1}^{\prime
\prime },$ $S_{2}^{\prime \prime }$. Heads of a coin is associated (as it is
in the three-coin setup) with $+1$ and tails with $-1$, and we denote the
outcome of Alice's, Bob's, and Chris' coins as $\pi _{A},$ $\pi _{B},$ and $%
\pi _{C}$, respectively.

Alice's outcome of $\pi _{A}=+1$ or $-1$, whether she goes for the $S_{1}$%
-coin or the $S_{2}$-coin in a run, is independent of Bob's outcome of $\pi
_{B}=+1$ or $-1$ as well as whether he goes for the $S_{1}^{\prime }$-coin
or the $S_{2}^{\prime }$-coin in the same run. Also, both of these outcomes
are independent of Chris' outcome of $\pi _{C}=+1$ or $-1$ as well as
whether he goes for the $S_{1}^{\prime \prime }$-coin or the $S_{2}^{\prime
\prime }$-coin in the same run.

The associated probabilities are, therefore, factorizable in the sense that
the probability for a triplet of outcomes can be expressed as the product of
probability for each outcome separately. Mathematically, this is expressed
by writing joint probabilities as the arithmetic product of their respective
marginals, i.e.

\begin{equation}
\Pr (\pi _{A},\pi _{B},\pi _{C};S_{1,2},S_{1,2}^{\prime },S_{1,2}^{\prime
\prime })=\Pr (\pi _{A},S_{1,2})\Pr (\pi _{B},S_{1,2}^{\prime })\Pr (\pi
_{C},S_{1,2}^{\prime \prime })
\end{equation}%
where, for example, Bob can set $S_{1,2}^{\prime }$ at $S_{1}^{\prime }$ or
at $S_{2}^{\prime }$ and the probability $\Pr (\pi _{A},\pi _{B},\pi
_{C};S_{2},S_{1}^{\prime },S_{2}^{\prime \prime })$ factorizes to $\Pr (\pi
_{A};S_{2})\Pr (\pi _{B};S_{1}^{\prime })\Pr (\pi _{C};S_{2}^{\prime \prime
})$.

As $S_{1},$ $S_{1}^{\prime },$ $S_{1}^{\prime \prime }$ are Alice's, Bob's,
and Chris' first coins, respectively, we assign probabilities $r,r^{\prime
},r^{\prime \prime }\in \lbrack 0,1]$ by defining $r=\Pr (+1;S_{1})$, $%
r^{\prime }=\Pr (+1;S_{1}^{\prime })$, and $r^{\prime \prime }=\Pr
(+1;S_{1}^{\prime \prime })$. Namely, $r$ is the probability of getting head
for (Alice's) $S_{1}$-coin; $r^{\prime }$ is the probability of getting head
for (Bob's) $S_{1}^{\prime }$-coin; and $r^{\prime \prime }$ is the
probability of getting head for (Chris') $S_{1}^{\prime \prime }$-coin.

Similarly, we assign probabilities $s,s^{\prime },s^{\prime \prime }\in
\lbrack 0,1]$ to $S_{2},$ $S_{2}^{\prime },$ $S_{2}^{\prime \prime }$ that
are Alice's, Bob's, and Chris' second coins, respectively: $s=\Pr (+1;S_{2})$%
, $s^{\prime }=\Pr (+1;S_{2}^{\prime })$, $s^{\prime \prime }=\Pr
(+1;S_{2}^{\prime \prime })$. Namely, $s$ is the probability of getting head
for (Alice's) $S_{2}$-coin; $s^{\prime }$ is the probability of getting head
for (Bob's) $S_{2}^{\prime }$-coin; and $s^{\prime \prime }$ is the
probability of getting head for (Chris') $S_{2}^{\prime \prime }$-coin.
Factorizability, then, for example, allows us to write $\Pr
(+1,-1,-1;S_{2},S_{1}^{\prime },S_{2}^{\prime \prime })=s(1-r^{\prime
})(1-s^{\prime \prime })$.

\subsubsection{Payoff relations and the Nash equilibria}

Given how we have defined a `pure strategy' in the six-coin setup the
players' pure-strategy payoff relations can now be written as

\begin{equation}
\begin{array}{l}
\Pi _{A,B,C}(S_{1},S_{1}^{\prime },S_{1}^{\prime \prime })=(\alpha ,\alpha
,\alpha )rr^{\prime }r^{\prime \prime }+(\delta ,\beta ,\delta
)r(1-r^{\prime })r^{\prime \prime }+(\delta ,\delta ,\beta )rr^{\prime
}(1-r^{\prime \prime })+ \\ 
(\epsilon ,\theta ,\theta )r(1-r^{\prime })(1-r^{\prime \prime })+(\beta
,\delta ,\delta )(1-r)r^{\prime }r^{\prime \prime }+(\theta ,\theta
,\epsilon )(1-r)(1-r^{\prime })r^{\prime \prime }+ \\ 
(\theta ,\epsilon ,\theta )(1-r)r^{\prime }(1-r^{\prime \prime })+(\omega
,\omega ,\omega )(1-r)(1-r^{\prime })(1-r^{\prime \prime });%
\end{array}
\label{6coin-1}
\end{equation}

\begin{equation}
\begin{array}{l}
\Pi _{A,B,C}(S_{2},S_{1}^{\prime },S_{1}^{\prime \prime })=(\alpha ,\alpha
,\alpha )sr^{\prime }r^{\prime \prime }+(\delta ,\beta ,\delta
)s(1-r^{\prime })r^{\prime \prime }+(\delta ,\delta ,\beta )sr^{\prime
}(1-r^{\prime \prime })+ \\ 
(\epsilon ,\theta ,\theta )s(1-r^{\prime })(1-r^{\prime \prime })+(\beta
,\delta ,\delta )(1-s)r^{\prime }r^{\prime \prime }+(\theta ,\theta
,\epsilon )(1-s)(1-r^{\prime })r^{\prime \prime }+ \\ 
(\theta ,\epsilon ,\theta )(1-s)r^{\prime }(1-r^{\prime \prime })+(\omega
,\omega ,\omega )(1-s)(1-r^{\prime })(1-r^{\prime \prime });%
\end{array}
\label{6coin-2}
\end{equation}

\begin{equation}
\begin{array}{l}
\Pi _{A,B,C}(S_{1},S_{2}^{\prime },S_{1}^{\prime \prime })=(\alpha ,\alpha
,\alpha )rs^{\prime }r^{\prime \prime }+(\delta ,\beta ,\delta
)r(1-s^{\prime })r^{\prime \prime }+(\delta ,\delta ,\beta )rs^{\prime
}(1-r^{\prime \prime })+ \\ 
(\epsilon ,\theta ,\theta )r(1-s^{\prime })(1-r^{\prime \prime })+(\beta
,\delta ,\delta )(1-r)s^{\prime }r^{\prime \prime }+(\theta ,\theta
,\epsilon )(1-r)(1-s^{\prime })r^{\prime \prime }+ \\ 
(\theta ,\epsilon ,\theta )(1-r)s^{\prime }(1-r^{\prime \prime })+(\omega
,\omega ,\omega )(1-r)(1-s^{\prime })(1-r^{\prime \prime });%
\end{array}
\label{6coin-3}
\end{equation}

\begin{equation}
\begin{array}{l}
\Pi _{A,B,C}(S_{1},S_{1}^{\prime },S_{2}^{\prime \prime })=(\alpha ,\alpha
,\alpha )rr^{\prime }s^{\prime \prime }+(\delta ,\beta ,\delta
)r(1-r^{\prime })s^{\prime \prime }+(\delta ,\delta ,\beta )rr^{\prime
}(1-s^{\prime \prime })+ \\ 
(\epsilon ,\theta ,\theta )r(1-r^{\prime })(1-s^{\prime \prime })+(\beta
,\delta ,\delta )(1-r)r^{\prime }s^{\prime \prime }+(\theta ,\theta
,\epsilon )(1-r)(1-r^{\prime })s^{\prime \prime }+ \\ 
(\theta ,\epsilon ,\theta )(1-r)r^{\prime }(1-s^{\prime \prime })+(\omega
,\omega ,\omega )(1-r)(1-r^{\prime })(1-s^{\prime \prime });%
\end{array}
\label{6coin-4}
\end{equation}

\begin{equation}
\begin{array}{l}
\Pi _{A,B,C}(S_{1},S_{2}^{\prime },S_{2}^{\prime \prime })=(\alpha ,\alpha
,\alpha )rs^{\prime }s^{\prime \prime }+(\delta ,\beta ,\delta
)r(1-s^{\prime })s^{\prime \prime }+(\delta ,\delta ,\beta )rs^{\prime
}(1-s^{\prime \prime })+ \\ 
(\epsilon ,\theta ,\theta )r(1-s^{\prime })(1-s^{\prime \prime })+(\beta
,\delta ,\delta )(1-r)s^{\prime }s^{\prime \prime }+(\theta ,\theta
,\epsilon )(1-r)(1-s^{\prime })s^{\prime \prime }+ \\ 
(\theta ,\epsilon ,\theta )(1-r)s^{\prime }(1-s^{\prime \prime })+(\omega
,\omega ,\omega )(1-r)(1-s^{\prime })(1-s^{\prime \prime });%
\end{array}
\label{6coin-5}
\end{equation}

\begin{equation}
\begin{array}{l}
\Pi _{A,B,C}(S_{2},S_{1}^{\prime },S_{2}^{\prime \prime })=(\alpha ,\alpha
,\alpha )sr^{\prime }s^{\prime \prime }+(\delta ,\beta ,\delta
)s(1-r^{\prime })s^{\prime \prime }+(\delta ,\delta ,\beta )sr^{\prime
}(1-s^{\prime \prime })+ \\ 
(\epsilon ,\theta ,\theta )s(1-r^{\prime })(1-s^{\prime \prime })+(\beta
,\delta ,\delta )(1-s)r^{\prime }s^{\prime \prime }+(\theta ,\theta
,\epsilon )(1-s)(1-r^{\prime })s^{\prime \prime }+ \\ 
(\theta ,\epsilon ,\theta )(1-s)r^{\prime }(1-s^{\prime \prime })+(\omega
,\omega ,\omega )(1-s)(1-r^{\prime })(1-s^{\prime \prime });%
\end{array}
\label{6coin-6}
\end{equation}

\begin{equation}
\begin{array}{l}
\Pi _{A,B,C}(S_{2},S_{2}^{\prime },S_{1}^{\prime \prime })=(\alpha ,\alpha
,\alpha )ss^{\prime }r^{\prime \prime }+(\delta ,\beta ,\delta
)s(1-s^{\prime })r^{\prime \prime }+(\delta ,\delta ,\beta )ss^{\prime
}(1-r^{\prime \prime })+ \\ 
(\epsilon ,\theta ,\theta )s(1-s^{\prime })(1-r^{\prime \prime })+(\beta
,\delta ,\delta )(1-s)s^{\prime }r^{\prime \prime }+(\theta ,\theta
,\epsilon )(1-s)(1-s^{\prime })r^{\prime \prime }+ \\ 
(\theta ,\epsilon ,\theta )(1-s)s^{\prime }(1-r^{\prime \prime })+(\omega
,\omega ,\omega )(1-s)(1-s^{\prime })(1-r^{\prime \prime });%
\end{array}
\label{6coin-7}
\end{equation}

\begin{equation}
\begin{array}{l}
\Pi _{A,B,C}(S_{2},S_{2}^{\prime },S_{2}^{\prime \prime })=(\alpha ,\alpha
,\alpha )ss^{\prime }s^{\prime \prime }+(\delta ,\beta ,\delta
)s(1-s^{\prime })s^{\prime \prime }+(\delta ,\delta ,\beta )ss^{\prime
}(1-s^{\prime \prime })+ \\ 
(\epsilon ,\theta ,\theta )s(1-s^{\prime })(1-s^{\prime \prime })+(\beta
,\delta ,\delta )(1-s)s^{\prime }s^{\prime \prime }+(\theta ,\theta
,\epsilon )(1-s)(1-s^{\prime })s^{\prime \prime }+ \\ 
(\theta ,\epsilon ,\theta )(1-s)s^{\prime }(1-s^{\prime \prime })+(\omega
,\omega ,\omega )(1-s)(1-s^{\prime })(1-s^{\prime \prime });%
\end{array}
\label{6coin-8}
\end{equation}%
where, on right side of each equation, the three constants in brackets
correspond to the players Alice, Bob, and Chris respectively. We point out
that Eqs.~(\ref{6coin-1}-\ref{6coin-8}) are pure-strategy payoff relations,
as they represent the situation when each of the three players chooses the
same coin for all runs. For example, referring to Eq.~(\ref{6coin-3}), $\Pi
_{A}(S_{1},S_{2}^{\prime },S_{1}^{\prime \prime })$ is Alice's payoff when
she goes for the $S_{1}$-coin, Bob goes for $S_{2}^{\prime }$-coin, and
Chris goes for $S_{1}^{\prime \prime }$-coin for all runs.

A mixed-strategy game, in the six-coin setup, corresponds to when, over a
large number of runs of the game, a player chooses one of the two available
coins with some probability. Let $x,y,$ and $z$ be the probabilities with
which Alice, Bob, and Chris, respectively, choose the coins $%
S_{1},S_{1}^{\prime },$ and $S_{1}^{\prime \prime }$, respectively. The
players' six-coin mixed-strategy payoff relations then read,

\begin{equation}
\begin{array}{l}
\Pi _{A,B,C}(x,y,z)=xyz\Pi _{A,B,C}(S_{1},S_{1}^{\prime },S_{1}^{\prime
\prime })+x(1-y)z\Pi _{A,B,C}(S_{1},S_{2}^{\prime },S_{1}^{\prime \prime })+
\\ 
xy(1-z)\Pi _{A,B,C}(S_{1},S_{1}^{\prime },S_{2}^{\prime \prime
})+x(1-y)(1-z)\Pi _{A,B,C}(S_{1},S_{2}^{\prime },S_{2}^{\prime \prime })+ \\ 
(1-x)yz\Pi _{A,B,C}(S_{2},S_{1}^{\prime },S_{1}^{\prime \prime
})+(1-x)(1-y)z\Pi _{A,B,C}(S_{2},S_{2}^{\prime },S_{1}^{\prime \prime })+ \\ 
(1-x)y(1-z)\Pi _{A,B,C}(S_{2},S_{1}^{\prime },S_{2}^{\prime \prime
})+(1-x)(1-y)(1-z)\Pi _{A,B,C}(S_{2},S_{2}^{\prime },S_{2}^{\prime \prime }).%
\end{array}
\label{6-coin mixed-strategy payoffs}
\end{equation}%
Notice that the right side of Eq.~(\ref{6-coin mixed-strategy payoffs})
contains expressions that are given by Eqs.~(\ref{6coin-1}-\ref{6coin-8}).

Six-coin mixed-strategy payoff relations (\ref{6-coin mixed-strategy payoffs}%
) are mathematically identical to the three-coin mixed-strategy payoff
relations (\ref{3-coin mixed-strategy payoffs}). However, these equations
are to be interpreted differently as the definitions of what constitutes a
strategy in three- and six-coin setups are different. In (\ref{3-coin
mixed-strategy payoffs}) the numbers $x,$ $y,$ and $z$ are the probabilities
with which Alice, Bob, and Chris, respectively, flip the coin that s/he
receives. Whereas in (\ref{6-coin mixed-strategy payoffs}) the numbers $x,$ $%
y,$ and $z$ are the probabilities with which, over repeated runs, Alice,
Bob, and Chris choose the $S_{1}$-coin, the $S_{1}^{\prime }$-coin, and the $%
S_{1}^{\prime \prime }$-coin, respectively.

Referring to (\ref{6-coin mixed-strategy payoffs}), the triplet $(x^{\star
},y^{\star },z^{\star })$ becomes a NE when the following inequalities hold%
\begin{equation}
\begin{array}{l}
\Pi _{A}(x^{\star },y^{\star },z^{\star })-\Pi _{A}(x,y^{\star },z^{\star
})\geqslant 0, \\ 
\Pi _{B}(x^{\star },y^{\star },z^{\star })-\Pi _{B}(x^{\star },y,z^{\star
})\geqslant 0, \\ 
\Pi _{C}(x^{\star },y^{\star },z^{\star })-\Pi _{B}(x^{\star },y^{\star
},z)\geqslant 0,%
\end{array}
\label{6-coin NE}
\end{equation}%
which, though being mathematically identical to (\ref{3-coin NE}), refers to
the relations (\ref{6-coin mixed-strategy payoffs}) and Eqs.~(\ref{6coin-1}-%
\ref{6coin-8}). Also, these inequalities are to be interpreted in terms of
how a strategy is defined in the six-coin setup.

\subsection{Playing Prisoner's Dilemma}

In the following we use the three-coin and the six-coin setups to play PD.

\subsubsection{With three-coin setup}

In three-coin setup, the inequalities (\ref{3-coin NE}) give

\begin{equation}
\begin{array}{l}
\Pi _{A}(x^{\star },y^{\star },z^{\star })-\Pi _{A}(x,y^{\star },z^{\star
})\geqslant 0, \\ 
\Pi _{B}(x^{\star },y^{\star },z^{\star })-\Pi _{B}(x^{\star },y,z^{\star
})\geqslant 0, \\ 
\Pi _{C}(x^{\star },y^{\star },z^{\star })-\Pi _{B}(x^{\star },y^{\star
},z)\geqslant 0,%
\end{array}%
\end{equation}%
and $(x^{\star },y^{\star },z^{\star })=(0,0,0)$ comes out as a unique NE at
which players' payoffs are $\Pi _{A}(0,0,0)=\omega ,$ $\Pi
_{B}(0,0,0)=\omega $, and $\Pi _{C}(0,0,0)=\omega .$

\subsubsection{With six-coin setup}

The NE conditions (\ref{6-coin NE}) are evaluated using the payoff relations
(\ref{6-coin mixed-strategy payoffs}) and the Eqs.~(\ref{6coin-1}-\ref%
{6coin-8}). We consider the case when $(s,s^{\prime },s^{\prime \prime
})=(0,0,0)$ i.e. the probabilities of getting head for each player's second
coin is zero. This reduces the NE conditions (\ref{6-coin NE}) to

\begin{equation}
\begin{array}{l}
(x^{\star }-x)\left\{ y^{\star }z^{\star }(rr^{\prime }r^{\prime \prime
})\Delta _{1}+r(z^{\star }r^{\prime \prime }+y^{\star }r^{\prime })\Delta
_{2}+r\Delta _{3}\right\} \geq 0, \\ 
(y^{\star }-y)\left\{ x^{\star }z^{\star }(rr^{\prime }r^{\prime \prime
})\Delta _{1}+r^{\prime }(z^{\star }r^{\prime \prime }+x^{\star }r)\Delta
_{2}+r^{\prime }\Delta _{3}\right\} \geq 0, \\ 
(z^{\star }-z)\left\{ x^{\star }y^{\star }(rr^{\prime }r^{\prime \prime
})\Delta _{1}+r^{\prime \prime }(y^{\star }r^{\prime }+x^{\star }r)\Delta
_{2}+r^{\prime \prime }\Delta _{3}\right\} \geq 0,%
\end{array}
\label{NE-Six-Coin PD}
\end{equation}%
where $\Delta _{1}=(\alpha -\beta -2\delta +2\theta +\epsilon -\omega ),$ $%
\Delta _{2}=(\delta -\epsilon -\theta +\omega ),$ and $\Delta _{3}=(\epsilon
-\omega )$. For PD we have $\Delta _{3}<0$, it then follows from (\ref%
{NE-Six-Coin PD}) that, when $(s,s^{\prime },s^{\prime \prime })=(0,0,0)$,
the strategy triplet $(x^{\star },y^{\star },z^{\star })=(0,0,0)$, as is
defined in the six-coin setup, comes out to be a unique NE in the
three-player symmetric game of PD. In other words, this is described by
saying that the triplet $(x^{\star },y^{\star },z^{\star })=(0,0,0)$ emerges
as a unique NE when the probabilities of getting a head for each player's
second coin is zero and, along with this, that the joint probabilities are
factorizable.

Now the crucial step follows: Requiring coins to satisfy the constraint $%
(s,s^{\prime },s^{\prime \prime })=(0,0,0)$ can also be translated in terms
of constraints on the joint probabilities associated to the six coins. To
find these constraints we identify the $64$ joint probabilities $%
p_{1},p_{2},...,p_{64}$ that can be defined for six coins, and are given as
follows:

\begin{equation}
\begin{array}{l}
p_{1}=rr^{\prime }r^{\prime \prime }, \\ 
p_{2}=r(1-r^{\prime })r^{\prime \prime }, \\ 
p_{3}=rr^{\prime }(1-r^{\prime \prime }), \\ 
p_{4}=r(1-r^{\prime })(1-r^{\prime \prime }),%
\end{array}%
\begin{array}{l}
p_{5}=(1-r)r^{\prime }r^{\prime \prime }, \\ 
p_{6}=(1-r)(1-r^{\prime })r^{\prime \prime }, \\ 
p_{7}=(1-r)r^{\prime }(1-r^{\prime \prime }), \\ 
p_{8}=(1-r)(1-r^{\prime })(1-r^{\prime \prime }),%
\end{array}
\label{6Coin probs 1}
\end{equation}

\begin{equation}
\begin{array}{l}
p_{9}=sr^{\prime }r^{\prime \prime }, \\ 
p_{10}=s(1-r^{\prime })r^{\prime \prime }, \\ 
p_{11}=sr^{\prime }(1-r^{\prime \prime }), \\ 
p_{12}=s(1-r^{\prime })(1-r^{\prime \prime }),%
\end{array}%
\begin{array}{l}
p_{13}=(1-s)r^{\prime }r^{\prime \prime }, \\ 
p_{14}=(1-s)(1-r^{\prime })r^{\prime \prime }, \\ 
p_{15}=(1-s)r^{\prime }(1-r^{\prime \prime }), \\ 
p_{16}=(1-s)(1-r^{\prime })(1-r^{\prime \prime }),%
\end{array}
\label{6Coin probs 2}
\end{equation}

\begin{equation}
\begin{array}{l}
p_{17}=rs^{\prime }r^{\prime \prime }, \\ 
p_{18}=r(1-s^{\prime })r^{\prime \prime }, \\ 
p_{19}=rs^{\prime }(1-r^{\prime \prime }), \\ 
p_{20}=r(1-s^{\prime })(1-r^{\prime \prime }),%
\end{array}%
\begin{array}{l}
p_{21}=(1-r)s^{\prime }r^{\prime \prime }, \\ 
p_{22}=(1-r)(1-s^{\prime })r^{\prime \prime }, \\ 
p_{23}=(1-r)s^{\prime }(1-r^{\prime \prime }), \\ 
p_{24}=(1-r)(1-s^{\prime })(1-r^{\prime \prime }),%
\end{array}
\label{6Coin probs 3}
\end{equation}

\begin{equation}
\begin{array}{l}
p_{25}=rr^{\prime }s^{\prime \prime }, \\ 
p_{26}=r(1-r^{\prime })s^{\prime \prime }, \\ 
p_{27}=rr^{\prime }(1-s^{\prime \prime }), \\ 
p_{28}=r(1-r^{\prime })(1-s^{\prime \prime }),%
\end{array}%
\begin{array}{l}
p_{29}=(1-r)r^{\prime }s^{\prime \prime }, \\ 
p_{30}=(1-r)(1-r^{\prime })s^{\prime \prime }, \\ 
p_{31}=(1-r)r^{\prime }(1-s^{\prime \prime }), \\ 
p_{32}=(1-r)(1-r^{\prime })(1-s^{\prime \prime }),%
\end{array}
\label{6Coin probs 4}
\end{equation}

\begin{equation}
\begin{array}{l}
p_{33}=rs^{\prime }s^{\prime \prime }, \\ 
p_{34}=r(1-s^{\prime })s^{\prime \prime }, \\ 
p_{35}=rs^{\prime }(1-s^{\prime \prime }), \\ 
p_{36}=r(1-s^{\prime })(1-s^{\prime \prime }),%
\end{array}%
\begin{array}{l}
p_{37}=(1-r)s^{\prime }s^{\prime \prime }, \\ 
p_{38}=(1-r)(1-s^{\prime })s^{\prime \prime }, \\ 
p_{39}=(1-r)s^{\prime }(1-s^{\prime \prime }), \\ 
p_{40}=(1-r)(1-s^{\prime })(1-s^{\prime \prime }),%
\end{array}
\label{6Coin probs 5}
\end{equation}

\begin{equation}
\begin{array}{l}
p_{41}=sr^{\prime }s^{\prime \prime }, \\ 
p_{42}=s(1-r^{\prime })s^{\prime \prime }, \\ 
p_{43}=sr^{\prime }(1-s^{\prime \prime }), \\ 
p_{44}=s(1-r^{\prime })(1-s^{\prime \prime }),%
\end{array}%
\begin{array}{l}
p_{45}=(1-s)r^{\prime }s^{\prime \prime }, \\ 
p_{46}=(1-s)(1-r^{\prime })s^{\prime \prime }, \\ 
p_{47}=(1-s)r^{\prime }(1-s^{\prime \prime }), \\ 
p_{48}=(1-s)(1-r^{\prime })(1-s^{\prime \prime }),%
\end{array}
\label{6Coin probs 6}
\end{equation}

\begin{equation}
\begin{array}{l}
p_{49}=ss^{\prime }r^{\prime \prime }, \\ 
p_{50}=s(1-s^{\prime })r^{\prime \prime }, \\ 
p_{51}=ss^{\prime }(1-r^{\prime \prime }), \\ 
p_{52}=s(1-s^{\prime })(1-r^{\prime \prime }),%
\end{array}%
\begin{array}{l}
p_{53}=(1-s)s^{\prime }r^{\prime \prime }, \\ 
p_{54}=(1-s)(1-s^{\prime })r^{\prime \prime }, \\ 
p_{55}=(1-s)s^{\prime }(1-r^{\prime \prime }), \\ 
p_{56}=(1-s)(1-s^{\prime })(1-r^{\prime \prime }),%
\end{array}
\label{6Coin probs 7}
\end{equation}

\begin{equation}
\begin{array}{l}
p_{57}=ss^{\prime }s^{\prime \prime }, \\ 
p_{58}=s(1-s^{\prime })s^{\prime \prime }, \\ 
p_{59}=ss^{\prime }(1-s^{\prime \prime }), \\ 
p_{60}=s(1-s^{\prime })(1-s^{\prime \prime }),%
\end{array}%
\begin{array}{l}
p_{61}=(1-s)s^{\prime }s^{\prime \prime }, \\ 
p_{62}=(1-s)(1-s^{\prime })s^{\prime \prime }, \\ 
p_{63}=(1-s)s^{\prime }(1-s^{\prime \prime }), \\ 
p_{64}=(1-s)(1-s^{\prime })(1-s^{\prime \prime }).%
\end{array}
\label{6Coin probs 8}
\end{equation}%
With these definitions the payoff relations (\ref{6coin-1}-\ref{6coin-8})
are re-expressed as

\begin{equation}
\begin{array}{l}
\Pi _{A,B,C}(S_{1},S_{1}^{\prime },S_{1}^{\prime \prime })=(\alpha ,\alpha
,\alpha )p_{1}+(\delta ,\beta ,\delta )p_{2}+(\delta ,\delta ,\beta )p_{3}+
\\ 
(\epsilon ,\theta ,\theta )p_{4}+(\beta ,\delta ,\delta )p_{5}+(\theta
,\theta ,\epsilon )p_{6}+(\theta ,\epsilon ,\theta )p_{7}+(\omega ,\omega
,\omega )p_{8};%
\end{array}
\label{6coin-1A}
\end{equation}

\begin{equation}
\begin{array}{l}
\Pi _{A,B,C}(S_{2},S_{1}^{\prime },S_{1}^{\prime \prime })=(\alpha ,\alpha
,\alpha )p_{9}+(\delta ,\beta ,\delta )p_{10}+(\delta ,\delta ,\beta )p_{11}+
\\ 
(\epsilon ,\theta ,\theta )p_{12}+(\beta ,\delta ,\delta )p_{13}+(\theta
,\theta ,\epsilon )p_{14}+(\theta ,\epsilon ,\theta )p_{15}+(\omega ,\omega
,\omega )p_{16};%
\end{array}
\label{6coin-2A}
\end{equation}

\begin{equation}
\begin{array}{l}
\Pi _{A,B,C}(S_{1},S_{2}^{\prime },S_{1}^{\prime \prime })=(\alpha ,\alpha
,\alpha )p_{17}+(\delta ,\beta ,\delta )p_{18}+(\delta ,\delta ,\beta
)p_{19}+ \\ 
(\epsilon ,\theta ,\theta )p_{20}+(\beta ,\delta ,\delta )p_{21}+(\theta
,\theta ,\epsilon )p_{22}+(\theta ,\epsilon ,\theta )p_{23}+(\omega ,\omega
,\omega )p_{24};%
\end{array}
\label{6coin-3A}
\end{equation}

\begin{equation}
\begin{array}{l}
\Pi _{A,B,C}(S_{1},S_{1}^{\prime },S_{2}^{\prime \prime })=(\alpha ,\alpha
,\alpha )p_{25}+(\delta ,\beta ,\delta )p_{26}+(\delta ,\delta ,\beta
)p_{27}+ \\ 
(\epsilon ,\theta ,\theta )p_{28}+(\beta ,\delta ,\delta )p_{29}+(\theta
,\theta ,\epsilon )p_{30}+(\theta ,\epsilon ,\theta )p_{31}+(\omega ,\omega
,\omega )p_{32};%
\end{array}
\label{6coin-4A}
\end{equation}

\begin{equation}
\begin{array}{l}
\Pi _{A,B,C}(S_{1},S_{2}^{\prime },S_{2}^{\prime \prime })=(\alpha ,\alpha
,\alpha )p_{33}+(\delta ,\beta ,\delta )p_{34}+(\delta ,\delta ,\beta
)p_{35}+ \\ 
(\epsilon ,\theta ,\theta )p_{36}+(\beta ,\delta ,\delta )p_{37}+(\theta
,\theta ,\epsilon )p_{38}+(\theta ,\epsilon ,\theta )p_{39}+(\omega ,\omega
,\omega )p_{40};%
\end{array}
\label{6coin-5A}
\end{equation}

\begin{equation}
\begin{array}{l}
\Pi _{A,B,C}(S_{2},S_{1}^{\prime },S_{2}^{\prime \prime })=(\alpha ,\alpha
,\alpha )p_{41}+(\delta ,\beta ,\delta )p_{42}+(\delta ,\delta ,\beta
)p_{43}+ \\ 
(\epsilon ,\theta ,\theta )p_{44}+(\beta ,\delta ,\delta )p_{45}+(\theta
,\theta ,\epsilon )p_{46}+(\theta ,\epsilon ,\theta )p_{47}+(\omega ,\omega
,\omega )p_{48};%
\end{array}
\label{6coin-6A}
\end{equation}

\begin{equation}
\begin{array}{l}
\Pi _{A,B,C}(S_{2},S_{2}^{\prime },S_{1}^{\prime \prime })=(\alpha ,\alpha
,\alpha )p_{49}+(\delta ,\beta ,\delta )p_{50}+(\delta ,\delta ,\beta
)p_{51}+ \\ 
(\epsilon ,\theta ,\theta )p_{52}+(\beta ,\delta ,\delta )p_{53}+(\theta
,\theta ,\epsilon )p_{54}+(\theta ,\epsilon ,\theta )p_{55}+(\omega ,\omega
,\omega )p_{56};%
\end{array}
\label{6coin-7A}
\end{equation}

\begin{equation}
\begin{array}{l}
\Pi _{A,B,C}(S_{2},S_{2}^{\prime },S_{2}^{\prime \prime })=(\alpha ,\alpha
,\alpha )p_{57}+(\delta ,\beta ,\delta )p_{58}+(\delta ,\delta ,\beta
)p_{59}+ \\ 
(\epsilon ,\theta ,\theta )p_{60}+(\beta ,\delta ,\delta )p_{61}+(\theta
,\theta ,\epsilon )p_{62}+(\theta ,\epsilon ,\theta )p_{63}+(\omega ,\omega
,\omega )p_{64}.%
\end{array}
\label{6coin-8A}
\end{equation}%
Now, from the definitions (\ref{6Coin probs 1}-\ref{6Coin probs 8}),
requiring coins to satisfy the constraint $(s,s^{\prime },s^{\prime \prime
})=(0,0,0)$ makes a number of the joint probabilities vanish:

\begin{equation}
\begin{array}{l}
p_{9}=0,\text{ }p_{10}=0,\text{ }p_{11}=0,\text{ }p_{12}=0; \\ 
p_{17}=0,\text{ }p_{19}=0,\text{ }p_{21}=0,\text{ }p_{23}=0; \\ 
p_{25}=0,\text{ }p_{26}=0,\text{ }p_{29}=0,\text{ }p_{30}=0; \\ 
p_{33}=0,\text{ }p_{34}=0,\text{ }p_{35}=0,\text{ }p_{37}=0,\text{ }p_{38}=0,%
\text{ }p_{39}=0; \\ 
p_{41}=0,\text{ }p_{42}=0,\text{ }p_{43}=0,\text{ }p_{44}=0,\text{ }p_{45}=0,%
\text{ }p_{46}=0; \\ 
p_{49}=0,\text{ }p_{50}=0,\text{ }p_{51}=0,\text{ }p_{52}=0,\text{ }p_{53}=0,%
\text{ }p_{55}=0; \\ 
p_{57}=0,\text{ }p_{58}=0,\text{ }p_{59}=0,\text{ }p_{60}=0,\text{ }p_{61}=0,%
\text{ }p_{62}=0,\text{ }p_{63}=0,%
\end{array}
\label{Constraints on joint probs}
\end{equation}%
which, in turn, reduces the pure-strategy payoff relations (\ref{6coin-1A}-%
\ref{6coin-8A}) to

\begin{equation}
\begin{array}{l}
\Pi _{A,B,C}(S_{1},S_{1}^{\prime },S_{1}^{\prime \prime })=(\alpha ,\alpha
,\alpha )p_{1}+(\delta ,\beta ,\delta )p_{2}+(\delta ,\delta ,\beta )p_{3}+
\\ 
(\epsilon ,\theta ,\theta )p_{4}+(\beta ,\delta ,\delta )p_{5}+(\theta
,\theta ,\epsilon )p_{6}+(\theta ,\epsilon ,\theta )p_{7}+(\omega ,\omega
,\omega )p_{8}; \\ 
\Pi _{A,B,C}(S_{2},S_{1}^{\prime },S_{1}^{\prime \prime })=(\beta ,\delta
,\delta )p_{13}+(\theta ,\theta ,\epsilon )p_{14}+(\theta ,\epsilon ,\theta
)p_{15}+(\omega ,\omega ,\omega )p_{16}; \\ 
\Pi _{A,B,C}(S_{1},S_{2}^{\prime },S_{1}^{\prime \prime })=(\delta ,\beta
,\delta )p_{18}+(\epsilon ,\theta ,\theta )p_{20}+(\theta ,\theta ,\epsilon
)p_{22}+(\omega ,\omega ,\omega )p_{24}; \\ 
\Pi _{A,B,C}(S_{1},S_{1}^{\prime },S_{2}^{\prime \prime })=(\delta ,\delta
,\beta )p_{27}+(\epsilon ,\theta ,\theta )p_{28}+(\theta ,\epsilon ,\theta
)p_{31}+(\omega ,\omega ,\omega )p_{32}; \\ 
\Pi _{A,B,C}(S_{1},S_{2}^{\prime },S_{2}^{\prime \prime })=(\epsilon ,\theta
,\theta )p_{36}+(\omega ,\omega ,\omega )p_{40}; \\ 
\Pi _{A,B,C}(S_{2},S_{1}^{\prime },S_{2}^{\prime \prime })=(\theta ,\epsilon
,\theta )p_{47}+(\omega ,\omega ,\omega )p_{48}; \\ 
\Pi _{A,B,C}(S_{2},S_{2}^{\prime },S_{1}^{\prime \prime })=(\theta ,\theta
,\epsilon )p_{54}+(\omega ,\omega ,\omega )p_{56}; \\ 
\Pi _{A,B,C}(S_{2},S_{2}^{\prime },S_{2}^{\prime \prime })=(\omega ,\omega
,\omega )p_{64};%
\end{array}
\label{6coin-B}
\end{equation}%
where only those joint probabilities are left that can have non-zero
value(s). These (pure strategy) payoff relations ensure that when the joint
probabilities (involved in these expressions) become factorizable the
classical outcome of the game results.

\section{Three-player quantum games}

In the quantum game literature \cite%
{MeyerDavid,EWL,BenjaminHayden,MarinattoWeber,Johnson,Piotrowski,Du,Flitney,Cheon,Shimamura,Iqbal}%
, three-player games have been studied but it appears that they have
attracted less attention then the two-player games. This is understandable
as their analysis is often found to be significantly harder even in the
classical regime.

As mentioned in Section~\ref{Intro}, an interesting example of a
three-player quantum game was discussed by Vaidman \cite{Vaidman,Vaidman1}
who described the GHZ paradox \cite{GHZ} as a game among three players.
Vaidman constructed a game, and tailored its winning conditions, such that
the winning chances for a classical team of three players cannot exceed $%
75\% $. A team of quantum players, however, are able to win the game $100\%$
if they share a GHZ state.

An analysis of Vaidman's game shows that it is won $100\%$ by a team of
quantum players having access to a probabilistic physical system for which
the joint probabilities are non-factorizable in a way described by the
winning conditions of the game. These conditions are constructed such that a
set of non-factorizable joint probabilities generated by the GHZ state
results in the team always winning the game.

Although Vaidman's game demonstrates how the GHZ state can be helpful in
winning a game, by itself this quantum game does not present a quantization
scheme for a general three-player noncooperative game. This was achieved by
Benjamin and Hayden \cite{BenjaminHayden} who developed a multiplayer
extension of Eisert et al.'s quantization scheme \cite{EWL}, originally
proposed for two-player noncooperative games. Eisert et al.'s scheme is
widely considered to have led to the birth of the area of quantum games.

However, Vaidman's game offers an interesting situation, which motivates one
to ask what may happen to a generalized three-player noncooperative
symmetric game, when the participating players share a probabilistic system
for which joint probabilities are not factorizable. This is precisely the
question that we aim to address in the present paper.

\subsection{Three-player quantum games using EPR-Bohm setting}

We consider an EPR-Bohm setting for three spatially-separated observers and
use it to play a general three-player symmetric noncooperative game. This
setting can be described as follows:

a) Three observers, henceforth called the players Alice, Bob, and Chris, are
distantly located and are not able to communicate among themselves.

b) In a run, each player receives a particle and has to choose one out of
the two directions and to inform the referee of his/her choice (henceforth
called his/her pure strategy).

c) The referee, after receiving information on the three players' strategies
in a run (which are three directions) rotates his Stern-Gerlach type
detectors along these directions and makes a measurement using Pauli spin
observables.

d) The two directions available to each player correspond to the two kinds
of (non-commuting) measurements that can be performed by the referee in a
run.

e) The outcome of a measurement, in a run, along any one of the three chosen
directions is either $+1$ or $-1$.

f) Over a large number of runs, a player can play a mixed strategy when s/he
has a linear combination (with normalized and real coefficients) of choosing
between the two available directions as the outcomes of quantum measurements
for all runs are recorded by the referee.

\FRAME{ftbpFU}{2.2416in}{2.5901in}{0pt}{\Qcb{In the EPR-Bohm setting for
playing a three-player quantum game, players Alice, Bob, and Chris each
receive\ a particle, in a run, coming from a tripartite state. Each player
has to decide one between the two available directions in the run and has to
inform the referee of his/her choice. The referee makes a quantum
measurement along the three chosen directions with Pauli spin operators. The
players' payoff relations are made public at the start of the game and
depend on the directions the players choose over a large number of runs
(defining their strategies), the matrix of the game, and on the joint
probabilities of the measurement outcomes that the referee obtains.}}{\Qlb{%
Fig1}}{fig1.eps}{\special{language "Scientific Word";type
"GRAPHIC";maintain-aspect-ratio TRUE;display "USEDEF";valid_file "F";width
2.2416in;height 2.5901in;depth 0pt;original-width 5.4587in;original-height
3.2154in;cropleft "0.3346";croptop "0.8787";cropright "0.7381";cropbottom
"0.0860";filename '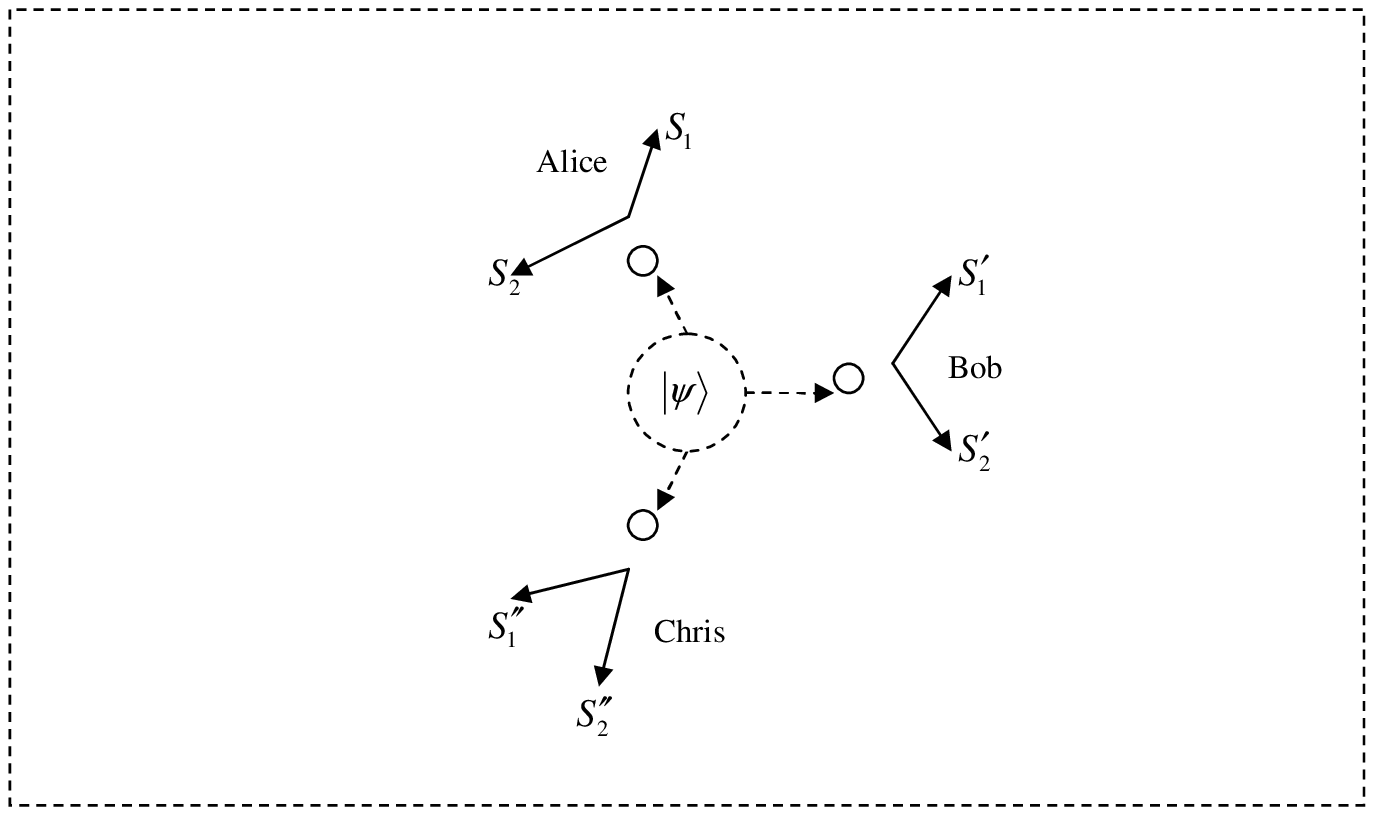';file-properties "XNPEU";}}

It can be noticed that in this EPR-Bohm setting, instead of coins, we can
let $S_{1}$ and $S_{2}$ to be Alice's two directions. Similarly, we can let $%
S_{1}^{\prime }$ and $S_{2}^{\prime }$ be Bob's two directions, and,
similarly, we can let $S_{1}^{\prime \prime }$ and $S_{2}^{\prime \prime }$
be Chris' two directions. As the outcome of a quantum measurement along one
of the three directions is $+1$ or $-1$ that we have associated to the Head
or Tail state of a coin, respectively.

Recall that in our six-coin setup, in a run, each player has to choose one
between two coins and to inform this choice to the referee. The referee,
after knowing which three coins the players have chosen in that run, tosses
them together and obtains a Head or Tail state for each.

Compare this to what happens in the EPR-Bohm setting in which, in a run, the
outcome of a quantum measurement is $+1$ or $-1$, along each one of three
directions chosen by the players. Note that players' strategies in the
EPR-Bohm setting are same as they are when they play classical strategies
using six coins.

This definition of strategy serves three purposes: a) it allows, in a
straight forward fashion, a quantum version of the three-player game from
the three-player classical game that is played using six coins b) as the
strategies remain identical in both the classical and the quantum games, it
seems that the Enk and Pike's argument \cite{EnkPike} is avoided\footnote{%
Essentially, this argument states that the process of allowing players
extended sets of strategies, in a quantum game setup that is based on Eisert
et al.'s scheme of playing a quantum game, can be equated to extending their
sets of pure strategies in the classical game.} c) it allows more direct
exploitation of the peculiar non-factorizable quantum mechanical
probabilities in terms of game-theoretic outcomes.

One may ask here about the quantum state(s) that are measured by the referee
after the players inform him/her of their strategies. In fact, apart from
the fact that the proposed setting uses three-partite quantum states, no
further restrictions are placed telling what should be the input quantum
states. That is, any pure or mixed three-partite states can be used to play
the quantum game.

\subsubsection{Joint probabilities in EPR-Bohm setting}

Using the above setting and noticing the notation for $64$ six-coin joint
probabilities in (\ref{6Coin probs 1}-\ref{6Coin probs 8}) convinces one to
introduce the same notation for $64$ joint probabilities corresponding to
the EPR-Bohm setting for playing a three-player quantum game:

\begin{equation}
\begin{array}{l}
p_{1}=\Pr (+1,+1,+1;S_{1},S_{1}^{\prime },S_{1}^{\prime \prime }), \\ 
p_{2}=\Pr (+1,-1,+1;S_{1},S_{1}^{\prime },S_{1}^{\prime \prime }), \\ 
p_{3}=\Pr (+1,+1,-1;S_{1},S_{1}^{\prime },S_{1}^{\prime \prime }), \\ 
p_{4}=\Pr (+1,-1,-1;S_{1},S_{1}^{\prime },S_{1}^{\prime \prime }),%
\end{array}%
\begin{array}{l}
p_{5}=\Pr (-1,+1,+1;S_{1},S_{1}^{\prime },S_{1}^{\prime \prime }), \\ 
p_{6}=\Pr (-1,-1,+1;S_{1},S_{1}^{\prime },S_{1}^{\prime \prime }), \\ 
p_{7}=\Pr (-1,+1,-1;S_{1},S_{1}^{\prime },S_{1}^{\prime \prime }), \\ 
p_{8}=\Pr (-1,-1,-1;S_{1},S_{1}^{\prime },S_{1}^{\prime \prime });%
\end{array}
\label{Joint Prob Def 1}
\end{equation}

\begin{equation}
\begin{array}{l}
p_{9}=\Pr (+1,+1,+1;S_{2},S_{1}^{\prime },S_{1}^{\prime \prime }), \\ 
p_{10}=\Pr (+1,-1,+1;S_{2},S_{1}^{\prime },S_{1}^{\prime \prime }), \\ 
p_{11}=\Pr (+1,+1,-1;S_{2},S_{1}^{\prime },S_{1}^{\prime \prime }), \\ 
p_{12}=\Pr (+1,-1,-1;S_{2},S_{1}^{\prime },S_{1}^{\prime \prime }),%
\end{array}%
\begin{array}{l}
p_{13}=\Pr (-1,+1,+1;S_{2},S_{1}^{\prime },S_{1}^{\prime \prime }), \\ 
p_{14}=\Pr (-1,-1,+1;S_{2},S_{1}^{\prime },S_{1}^{\prime \prime }), \\ 
p_{15}=\Pr (-1,+1,-1;S_{2},S_{1}^{\prime },S_{1}^{\prime \prime }), \\ 
p_{16}=\Pr (-1,-1,-1;S_{2},S_{1}^{\prime },S_{1}^{\prime \prime });%
\end{array}
\label{Joint Prob Def 2}
\end{equation}

\begin{equation}
\begin{array}{l}
p_{17}=\Pr (+1,+1,+1;S_{1},S_{2}^{\prime },S_{1}^{\prime \prime }), \\ 
p_{18}=\Pr (+1,-1,+1;S_{1},S_{2}^{\prime },S_{1}^{\prime \prime }), \\ 
p_{19}=\Pr (+1,+1,-1;S_{1},S_{2}^{\prime },S_{1}^{\prime \prime }), \\ 
p_{20}=\Pr (+1,-1,-1;S_{1},S_{2}^{\prime },S_{1}^{\prime \prime }),%
\end{array}%
\begin{array}{l}
p_{21}=\Pr (-1,+1,+1;S_{1},S_{2}^{\prime },S_{1}^{\prime \prime }), \\ 
p_{22}=\Pr (-1,-1,+1;S_{1},S_{2}^{\prime },S_{1}^{\prime \prime }), \\ 
p_{23}=\Pr (-1,+1,-1;S_{1},S_{2}^{\prime },S_{1}^{\prime \prime }), \\ 
p_{24}=\Pr (-1,-1,-1;S_{1},S_{2}^{\prime },S_{1}^{\prime \prime });%
\end{array}
\label{Joint Prob Def 3}
\end{equation}

\begin{equation}
\begin{array}{l}
p_{25}=\Pr (+1,+1,+1;S_{1},S_{1}^{\prime },S_{2}^{\prime \prime }), \\ 
p_{26}=\Pr (+1,-1,+1;S_{1},S_{1}^{\prime },S_{2}^{\prime \prime }), \\ 
p_{27}=\Pr (+1,+1,-1;S_{1},S_{1}^{\prime },S_{2}^{\prime \prime }), \\ 
p_{28}=\Pr (+1,-1,-1;S_{1},S_{1}^{\prime },S_{2}^{\prime \prime }),%
\end{array}%
\begin{array}{l}
p_{29}=\Pr (-1,+1,+1;S_{1},S_{1}^{\prime },S_{2}^{\prime \prime }), \\ 
p_{30}=\Pr (-1,-1,+1;S_{1},S_{1}^{\prime },S_{2}^{\prime \prime }), \\ 
p_{31}=\Pr (-1,+1,-1;S_{1},S_{1}^{\prime },S_{2}^{\prime \prime }), \\ 
p_{32}=\Pr (-1,-1,-1;S_{1},S_{1}^{\prime },S_{2}^{\prime \prime });%
\end{array}
\label{Joint Prob Def 4}
\end{equation}

\begin{equation}
\begin{array}{l}
p_{33}=\Pr (+1,+1,+1;S_{1},S_{2}^{\prime },S_{2}^{\prime \prime }), \\ 
p_{34}=\Pr (+1,-1,+1;S_{1},S_{2}^{\prime },S_{2}^{\prime \prime }), \\ 
p_{35}=\Pr (+1,+1,-1;S_{1},S_{2}^{\prime },S_{2}^{\prime \prime }), \\ 
p_{36}=\Pr (+1,-1,-1;S_{1},S_{2}^{\prime },S_{2}^{\prime \prime }),%
\end{array}%
\begin{array}{l}
p_{37}=\Pr (-1,+1,+1;S_{1},S_{2}^{\prime },S_{2}^{\prime \prime }), \\ 
p_{38}=\Pr (-1,-1,+1;S_{1},S_{2}^{\prime },S_{2}^{\prime \prime }), \\ 
p_{39}=\Pr (-1,+1,-1;S_{1},S_{2}^{\prime },S_{2}^{\prime \prime }), \\ 
p_{40}=\Pr (-1,-1,-1;S_{1},S_{2}^{\prime },S_{2}^{\prime \prime });%
\end{array}
\label{Joint Prob Def 5}
\end{equation}

\begin{equation}
\begin{array}{l}
p_{41}=\Pr (+1,+1,+1;S_{2},S_{1}^{\prime },S_{2}^{\prime \prime }), \\ 
p_{42}=\Pr (+1,-1,+1;S_{2},S_{1}^{\prime },S_{2}^{\prime \prime }), \\ 
p_{43}=\Pr (+1,+1,-1;S_{2},S_{1}^{\prime },S_{2}^{\prime \prime }), \\ 
p_{44}=\Pr (+1,-1,-1;S_{2},S_{1}^{\prime },S_{2}^{\prime \prime }),%
\end{array}%
\begin{array}{l}
p_{45}=\Pr (-1,+1,+1;S_{2},S_{1}^{\prime },S_{2}^{\prime \prime }), \\ 
p_{46}=\Pr (-1,-1,+1;S_{2},S_{1}^{\prime },S_{2}^{\prime \prime }), \\ 
p_{47}=\Pr (-1,+1,-1;S_{2},S_{1}^{\prime },S_{2}^{\prime \prime }), \\ 
p_{48}=\Pr (-1,-1,-1;S_{2},S_{1}^{\prime },S_{2}^{\prime \prime });%
\end{array}
\label{Joint Prob Def 6}
\end{equation}

\begin{equation}
\begin{array}{l}
p_{49}=\Pr (+1,+1,+1;S_{2},S_{2}^{\prime },S_{1}^{\prime \prime }), \\ 
p_{50}=\Pr (+1,-1,+1;S_{2},S_{2}^{\prime },S_{1}^{\prime \prime }), \\ 
p_{51}=\Pr (+1,+1,-1;S_{2},S_{2}^{\prime },S_{1}^{\prime \prime }), \\ 
p_{52}=\Pr (+1,-1,-1;S_{2},S_{2}^{\prime },S_{1}^{\prime \prime }),%
\end{array}%
\begin{array}{l}
p_{53}=\Pr (-1,+1,+1;S_{2},S_{2}^{\prime },S_{1}^{\prime \prime }), \\ 
p_{54}=\Pr (-1,-1,+1;S_{2},S_{2}^{\prime },S_{1}^{\prime \prime }), \\ 
p_{55}=\Pr (-1,+1,-1;S_{2},S_{2}^{\prime },S_{1}^{\prime \prime }), \\ 
p_{56}=\Pr (-1,-1,-1;S_{2},S_{2}^{\prime },S_{1}^{\prime \prime });%
\end{array}
\label{Joint Prob Def 7}
\end{equation}

\begin{equation}
\begin{array}{l}
p_{57}=\Pr (+1,+1,+1;S_{2},S_{2}^{\prime },S_{2}^{\prime \prime }), \\ 
p_{58}=\Pr (+1,-1,+1;S_{2},S_{2}^{\prime },S_{2}^{\prime \prime }), \\ 
p_{59}=\Pr (+1,+1,-1;S_{2},S_{2}^{\prime },S_{2}^{\prime \prime }), \\ 
p_{60}=\Pr (+1,-1,-1;S_{2},S_{2}^{\prime },S_{2}^{\prime \prime }),%
\end{array}%
\begin{array}{l}
p_{61}=\Pr (-1,+1,+1;S_{2},S_{2}^{\prime },S_{2}^{\prime \prime }), \\ 
p_{62}=\Pr (-1,-1,+1;S_{2},S_{2}^{\prime },S_{2}^{\prime \prime }), \\ 
p_{63}=\Pr (-1,+1,-1;S_{2},S_{2}^{\prime },S_{2}^{\prime \prime }), \\ 
p_{64}=\Pr (-1,-1,-1;S_{2},S_{2}^{\prime },S_{2}^{\prime \prime }).%
\end{array}
\label{Joint Prob Def 8}
\end{equation}%
Notice that for these coins the joint probabilities are factorizable and (%
\ref{Joint Prob Def 1}-\ref{Joint Prob Def 8}) reduce to (\ref{6Coin probs 1}%
-\ref{6Coin probs 8}).

\subsubsection{Constraints on joint probabilities}

We note that quantum mechanics imposes constraints on the joint
probabilities (\ref{Joint Prob Def 1}-\ref{Joint Prob Def 8}) that are
usually known \cite{Cereceda} as the normalization and the causal
communication constraints. Using the definitions (\ref{Joint Prob Def 1}-\ref%
{Joint Prob Def 8}) the constraints imposed by normalization are:

\begin{equation}
\begin{array}{l}
\sum\nolimits_{i=1}^{8}p_{i}=1,\text{ }\sum\nolimits_{i=9}^{16}p_{i}=1,\text{
}\sum\nolimits_{i=17}^{24}p_{i}=1,\text{ }\sum\nolimits_{i=25}^{32}p_{i}=1,
\\ 
\\ 
\sum\nolimits_{i=33}^{40}p_{i}=1,\text{ }\sum\nolimits_{i=41}^{48}p_{i}=1,%
\text{ }\sum\nolimits_{i=49}^{56}p_{i}=1,\text{ }\sum%
\nolimits_{i=57}^{64}p_{i}=1.%
\end{array}
\label{normalization}
\end{equation}

The first equation in (\ref{normalization}) with reference to the
definitions (\ref{Joint Prob Def 1}) of the joint probabilities $%
p_{1},p_{2},...p_{8}$, for example, describes the situation when Alice, Bob,
and Chris play $S_{1},$ $S_{1}^{\prime },$ and $S_{1}^{\prime \prime }$,
respectively, over all runs. In this case the probabilities $%
p_{1},p_{2},...p_{8}$ correspond to the outcomes $(+1,+1,+1),$ $(+1,-1,+1),$ 
$(+1,+1,-1),$ $(+1,-1,-1),$ $(-1,+1,+1),$ $(-1,-1,+1),$ $(-1,+1,-1),$ and $%
(-1,-1,-1)$, respectively. As these are the only possible outcomes for the
strategy triplet $(S_{1},$ $S_{1}^{\prime },$ $S_{1}^{\prime \prime })$, the
corresponding probabilities must add up to one. The remaining seven
equations in (\ref{normalization}) can be interpreted similarly.

Notice that in six-coin setup the triplet $(S_{1},$ $S_{1}^{\prime },$ $%
S_{1}^{\prime \prime })$ refers to the situation when Alice, Bob, and Chris
go for the $S_{1}$-coin, the $S_{1}^{\prime }$-coin, and the $S_{1}^{\prime
\prime }$-coin, respectively, for all runs. The corresponding eight coin
probabilities (\ref{6Coin probs 1}) are normalized and satisfy the first
equation in (\ref{normalization}). The same holds true for the remaining
coin probabilities (\ref{6Coin probs 2}-\ref{6Coin probs 8}) as they satisfy
the remaining equations in (\ref{normalization}).

Along with the constraints that normalization imposes on joint
probabilities, there are other constraints that are imposed by requirements
of causal communication. Often these constraints are referred \cite{Cereceda}
to as the condition of `parameter independence', `simple locality', `signal
locality', or `physical locality'. Essentially, they say, for example, that
the probability $\Pr^{A}(+1;S_{1})$ of Alice obtaining the outcome $+1$ when
she plays $S_{1}$ is not changed by Bob's choice for $S_{1}^{\prime }$ or $%
S_{2}^{\prime }$ and Chris' choice for $S_{1}^{\prime \prime }$ or $%
S_{2}^{\prime \prime }$. That is, the probability of a particular
measurement outcome on one part of the system is independent of which kind
of measurement is performed by the referee on the other part(s) \cite%
{Cereceda}. Causal communication constraints make it impossible for the
participating agents to acausually exchange the classical information.

The probability $\Pr^{A}(+1;S_{1})$ corresponds to when Alice chooses $S_{1}$
and the referee obtains the outcome $+1$ along $S_{1}$. Independence of
Bob's and Chris' choices requires that,

\begin{equation}
\Pr^{A}(+1;S_{1},S_{1}^{\prime },S_{1}^{\prime \prime
})=\Pr^{A}(+1;S_{1},S_{1}^{\prime },S_{2}^{\prime \prime
})=\Pr^{A}(+1;S_{1},S_{2}^{\prime },S_{1}^{\prime \prime
})=\Pr^{A}(+1;S_{1},S_{2}^{\prime },S_{2}^{\prime \prime }),
\end{equation}%
which can be expanded using (\ref{Joint Prob Def 1}-\ref{Joint Prob Def 8})
as

\begin{equation}
\begin{array}{cc}
\Pr (+1;S_{1}) & =\Pr (+1,+1,+1;S_{1},S_{1}^{\prime },S_{1}^{\prime \prime
})+\Pr (+1,+1,-1;S_{1},S_{1}^{\prime },S_{1}^{\prime \prime })+ \\ 
& \Pr (+1,-1,+1;S_{1},S_{1}^{\prime },S_{1}^{\prime \prime })+\Pr
(+1,-1,-1;S_{1},S_{1}^{\prime },S_{1}^{\prime \prime }) \\ 
& =\Pr (+1,+1,+1;S_{1},S_{1}^{\prime },S_{2}^{\prime \prime })+\Pr
(+1,+1,-1;S_{1},S_{1}^{\prime },S_{2}^{\prime \prime })+ \\ 
& \Pr (+1,-1,+1;S_{1},S_{1}^{\prime },S_{2}^{\prime \prime })+\Pr
(+1,-1,-1;S_{1},S_{1}^{\prime },S_{2}^{\prime \prime }) \\ 
& =\Pr (+1,+1,+1;S_{1},S_{2}^{\prime },S_{1}^{\prime \prime })+\Pr
(+1,+1,-1;S_{1},S_{2}^{\prime },S_{1}^{\prime \prime })+ \\ 
& \Pr (+1,-1,+1;S_{1},S_{2}^{\prime },S_{1}^{\prime \prime })+\Pr
(+1,-1,-1;S_{1},S_{2}^{\prime },S_{1}^{\prime \prime }) \\ 
& =\Pr (+1,+1,+1;S_{1},S_{2}^{\prime },S_{2}^{\prime \prime })+\Pr
(+1,+1,-1;S_{1},S_{2}^{\prime },S_{2}^{\prime \prime })+ \\ 
& \Pr (+1,-1,+1;S_{1},S_{2}^{\prime },S_{2}^{\prime \prime })+\Pr
(+1,-1,-1;S_{1},S_{2}^{\prime },S_{2}^{\prime \prime }).%
\end{array}
\label{causal communication}
\end{equation}%
This reads as,

\begin{equation}
p_{1}+p_{2}+p_{3}+p_{4}=p_{25}+p_{26}+p_{27}+p_{28}=p_{17}+p_{18}+p_{19}+p_{20}=p_{33}+p_{34}+p_{35}+p_{36}.
\label{locality-0}
\end{equation}

Constraints similar to (\ref{causal communication}) can be written for the
probabilities $\overset{A}{\Pr }(-1;S_{1}),$ $\overset{A}{\Pr }(+1;S_{2}),$ $%
\overset{A}{\Pr }(-1;S_{2})$ that correspond to Alice's choices; for the
probabilities $\overset{B}{\Pr }(+1;S_{1}^{\prime }),$ $\overset{B}{\Pr }%
(-1;S_{1}^{\prime }),$ $\overset{B}{\Pr }(+1;S_{2}^{\prime }),$ $\overset{B}{%
\Pr }(-1;S_{2}^{\prime })$ that correspond to Bob's choices; and for the
probabilities $\overset{C}{\Pr }(+1;S_{1}^{\prime \prime }),$ $\overset{C}{%
\Pr }(-1;S_{1}^{\prime \prime }),$ $\overset{C}{\Pr }(+1;S_{2}^{\prime
\prime }),$ $\overset{C}{\Pr }(-1;S_{2}^{\prime \prime })$ that correspond
to Chris' choices.

Using the definitions (\ref{Joint Prob Def 1}-\ref{Joint Prob Def 8}) for
Alice's choices these constraints read

\begin{equation}
\begin{array}{l}
p_{5}+p_{6}+p_{7}+p_{8}=p_{29}+p_{30}+p_{31}+p_{32}=p_{21}+p_{22}+p_{23}+p_{24}=p_{37}+p_{38}+p_{39}+p_{40},
\\ 
p_{9}+p_{10}+p_{11}+p_{12}=p_{41}+p_{42}+p_{43}+p_{44}=p_{49}+p_{50}+p_{51}+p_{52}=p_{57}+p_{58}+p_{59}+p_{60},
\\ 
p_{13}+p_{14}+p_{15}+p_{16}=p_{45}+p_{46}+p_{47}+p_{48}=p_{53}+p_{54}+p_{55}+p_{56}=p_{61}+p_{62}+p_{63}+p_{64},%
\end{array}
\label{locality-1}
\end{equation}%
and for Bob's choices they read

\begin{equation}
\begin{array}{l}
p_{1}+p_{3}+p_{5}+p_{7}=p_{25}+p_{27}+p_{29}+p_{31}=p_{9}+p_{11}+p_{13}+p_{15}=p_{41}+p_{43}+p_{45}+p_{47},
\\ 
p_{2}+p_{4}+p_{6}+p_{8}=p_{26}+p_{28}+p_{30}+p_{32}=p_{10}+p_{12}+p_{14}+p_{16}=p_{42}+p_{44}+p_{46}+p_{48},
\\ 
p_{17}+p_{19}+p_{21}+p_{23}=p_{33}+p_{35}+p_{37}+p_{39}=p_{49}+p_{51}+p_{53}+p_{55}=p_{57}+p_{59}+p_{61}+p_{63},
\\ 
p_{18}+p_{20}+p_{22}+p_{24}=p_{34}+p_{36}+p_{38}+p_{40}=p_{50}+p_{52}+p_{54}+p_{56}=p_{58}+p_{60}+p_{62}+p_{64},%
\end{array}
\label{locality-2}
\end{equation}%
and for Chris' choices they read

\begin{equation}
\begin{array}{l}
p_{1}+p_{2}+p_{5}+p_{6}=p_{17}+p_{18}+p_{21}+p_{22}=p_{9}+p_{10}+p_{13}+p_{14}=p_{49}+p_{50}+p_{53}+p_{54},
\\ 
p_{3}+p_{4}+p_{7}+p_{8}=p_{19}+p_{20}+p_{23}+p_{24}=p_{11}+p_{12}+p_{15}+p_{16}=p_{51}+p_{52}+p_{55}+p_{56},
\\ 
p_{25}+p_{26}+p_{29}+p_{30}=p_{33}+p_{34}+p_{37}+p_{38}=p_{41}+p_{42}+p_{45}+p_{46}=p_{57}+p_{58}+p_{61}+p_{62},
\\ 
p_{27}+p_{28}+p_{31}+p_{32}=p_{35}+p_{36}+p_{39}+p_{40}=p_{43}+p_{44}+p_{47}+p_{48}=p_{59}+p_{60}+p_{63}+p_{64}.%
\end{array}
\label{locality-3}
\end{equation}%
Notice that the coin probabilities (\ref{6Coin probs 1}-\ref{6Coin probs 8})
satisfy the causal communication constraints (\ref{locality-0}-\ref%
{locality-3}) along with the constraints (\ref{normalization}) imposed by
normalization.

However, in an EPR-Bohm setting involving three observers, there may emerge
a set of quantum mechanical joint probabilities (\ref{Joint Prob Def 1}-\ref%
{Joint Prob Def 8}) that cannot be expressed in the form of the joint
probabilities (\ref{6Coin probs 1}-\ref{6Coin probs 8}), which correspond to
six coins. In other words this means that for this set it is not possible to
find six numbers $r,s,r^{\prime },s^{\prime },r^{\prime \prime },s^{\prime
\prime }\in \lbrack 0,1]$ such that the set (\ref{Joint Prob Def 1}-\ref%
{Joint Prob Def 8}) can be reproduced from the coin probabilities using (\ref%
{6Coin probs 1}-\ref{6Coin probs 8}).

\subsection{Three-player quantum Prisoner's Dilemma}

The set of joint probabilities in the six-coin setup is factorizable in
terms of six probabilities $r,$ $s,$ $r^{\prime },$ $s^{\prime },$ $%
r^{\prime \prime },$ $s^{\prime \prime }$ and the constraints (\ref%
{Constraints on joint probs}) ensure that this set leads to the emergence of
the strategy triplet $(x^{\star },y^{\star },z^{\star })=(0,0,0)$ as a NE in
three-player symmetric game of PD. Whereas a set of quantum mechanical joint
probabilities that is obtained, for example, from three qubits in some
quantum state, is not necessarily factorizable.

To ensure that the classical game and its classical solution remains
embedded in the quantum game we require that the set of quantum mechanical
joint probabilities (\ref{Joint Prob Def 1}-\ref{Joint Prob Def 8}) also
satisfies the constraints (\ref{Constraints on joint probs}). This is
central to present argument as it guarantees that when the set (\ref{Joint
Prob Def 1}-\ref{Joint Prob Def 8}) is factorizable the strategy triplet $%
(D,D,D)$, also represented as $(x^{\star },y^{\star },z^{\star })=(0,0,0)$,
comes out to be the unique NE of the game. When the constraints (\ref%
{Constraints on joint probs}) are fulfilled, the relations (\ref%
{normalization}) describing normalization and the relations (\ref{locality-1}%
-\ref{locality-3}), which describe the causal communication constraint, are
reduced to

\begin{equation}
\begin{array}{l}
p_{1}+p_{2}+p_{3}+p_{4}+p_{5}+p_{6}+p_{7}+p_{8}=1, \\ 
p_{13}+p_{14}+p_{15}+p_{16}=1, \\ 
p_{18}+p_{20}+p_{22}+p_{24}=1, \\ 
p_{27}+p_{28}+p_{31}+p_{32}=1, \\ 
p_{36}+p_{40}=1,\text{ }p_{47}+p_{48}=1, \\ 
p_{54}+p_{56}=1,\text{ }p_{64}=1;%
\end{array}
\label{normalization-PD}
\end{equation}

\begin{equation}
\begin{array}{l}
p_{1}+p_{2}+p_{3}+p_{4}=p_{27}+p_{28}=p_{18}+p_{20}=p_{36}, \\ 
p_{5}+p_{6}+p_{7}+p_{8}=p_{31}+p_{32}=p_{22}+p_{24}=p_{40}, \\ 
p_{13}+p_{14}+p_{15}+p_{16}=1;%
\end{array}
\label{locality-1-PD}
\end{equation}

\begin{equation}
\begin{array}{l}
p_{1}+p_{3}+p_{5}+p_{7}=p_{27}+p_{31}=p_{13}+p_{15}=p_{47}, \\ 
p_{2}+p_{4}+p_{6}+p_{8}=p_{28}+p_{32}=p_{14}+p_{16}=p_{48}, \\ 
p_{18}+p_{20}+p_{22}+p_{24}=1;%
\end{array}
\label{locality-2-PD}
\end{equation}

\begin{equation}
\begin{array}{l}
p_{1}+p_{2}+p_{5}+p_{6}=p_{18}+p_{22}=p_{13}+p_{14}=p_{54}, \\ 
p_{3}+p_{4}+p_{7}+p_{8}=p_{20}+p_{24}=p_{15}+p_{16}=p_{56}, \\ 
p_{27}+p_{28}+p_{31}+p_{32}=1.%
\end{array}
\label{locality-3-PD}
\end{equation}

Requiring a set of $64$ joint probabilities to satisfy the constraints (\ref%
{Constraints on joint probs}) embeds the classical game within a quantum one
as when the set becomes factorizable the strategy triplet $(D,D,D)\equiv
(0,0,0)$ emerges as a unique NE. Now, for PD the most interesting situation
is when the strategy of Cooperation $(C)$, on the behalf of all the three
players, comes out as a NE of the game---as this is also the Pareto-optimal 
\cite{Rasmusen} solution of the game. In our notation, $(C,C,C)$ is given by 
$(x^{\star },y^{\star },z^{\star })=(1,1,1)$.

\subsubsection{Non-factorizable probabilities generating the pareto-optimal
Nash equilibrium}

We now explore whether the strategy profile $(C,C,C)$ can be a NE for a set
of non-factorizable, and thus quantum mechanical, joint probabilities. The
explicit expressions, which state the conditions (\ref{6-coin NE}) for the
strategy profile $(C,C,C)$ to be a NE in PD, can be found as follows. Use
the payoff relations (\ref{6-coin mixed-strategy payoffs}) and (\ref{6coin-B}%
) along with the constraints (\ref{normalization-PD}-\ref{locality-3-PD}) to
reduce the conditions (\ref{6-coin NE}) to

\begin{equation}
\begin{array}{l}
\left\{ p_{5}+(\alpha /\beta )p_{1}-p_{13}\right\} +(\theta /\beta )\left\{
p_{6}+p_{7}-p_{14}-p_{15}+(\delta /\theta )(p_{2}+p_{3})\right\} + \\ 
(\omega /\beta )\left\{ p_{8}-p_{16}+(\epsilon /\omega )p_{4}\right\} \geq 0;%
\end{array}
\label{(C,C,C)NE(A)}
\end{equation}

\begin{equation}
\begin{array}{l}
\left\{ p_{2}+(\alpha /\beta )p_{1}-p_{18}\right\} +(\theta /\beta )\left\{
p_{4}+p_{6}-p_{20}-p_{22}+(\delta /\theta )(p_{3}+p_{5})\right\} + \\ 
(\omega /\beta )\left\{ p_{8}-p_{24}+(\epsilon /\omega )p_{7}\right\} \geq 0;%
\end{array}
\label{(C,C,C)NE(B)}
\end{equation}

\begin{equation}
\begin{array}{l}
\left\{ p_{3}+(\alpha /\beta )p_{1}-p_{27}\right\} +(\theta /\beta )\left\{
p_{4}+p_{7}-p_{28}-p_{31}+(\delta /\theta )(p_{2}+p_{5})\right\} + \\ 
(\omega /\beta )\left\{ p_{8}-p_{32}+(\epsilon /\omega )p_{6}\right\} \geq 0;%
\end{array}
\label{(C,C,C)NE(C)}
\end{equation}%
where the definition (\ref{Generalized PD Definition}) of generalized
three-player PD requires that each one of the five quantities $\alpha /\beta
,$ $\theta /\beta ,$ $\delta /\theta ,$ $\omega /\beta ,$ $\epsilon /\omega $
are less than $1$.

With this our question, whether the strategy profile $(C,C,C)$ can exist as
a NE in a symmetric three-player quantum game of PD, is now re-expressed in
terms of finding five quantities $\alpha /\beta ,$ $\theta /\beta ,$ $\delta
/\theta ,$ $\omega /\beta ,$ $\epsilon /\omega $ (all less than $1$) and a
set of $64$ (non-factorizable) joint probabilities for which the conditions (%
\ref{normalization-PD}, \ref{locality-1-PD}-\ref{locality-3-PD}) as well as
the constraints (\ref{Constraints on joint probs}) hold. If the inequalities
(\ref{(C,C,C)NE(A)}-\ref{(C,C,C)NE(C)}) hold for these five quantities and
for the set of $64$ (non-factorizable) joint probabilities, it will show
that the strategy profile $(C,C,C)$ becomes a NE in the quantum game.

It turns out that, in fact, it is not difficult to find such five
quantities, all less than $1$, and a set of $64$ joint probabilities for
which all of the above requirements hold. Consider the following example.
Assign values $\alpha /\beta =9/10,$ $\theta /\beta =1/100,$ $\delta /\theta
=1/5,$ $\omega /\beta =1/100,$ and $\epsilon /\omega =9/10$ for which the
game becomes that of PD. Let the ten probabilities $p_{1},$ $p_{3},$ $p_{5},$
$p_{6},$ $p_{13},$ $p_{15},$ $p_{18},$ $p_{20},$ $p_{22},$ $p_{27}$ be
`independent' and assign values to them as $p_{1}=1/10,$ $p_{3}=13/100,$ $%
p_{5}=4/25,$ $p_{6}=1/10,$ $p_{13}=7/50,$ $p_{15}=2/5,$ $p_{18}=13/100,$ $%
p_{20}=1/4,$ $p_{22}=37/100,$ and $p_{27}=1/5$. Note that the constraints (%
\ref{Constraints on joint probs}) assign zero value to thirty seven
probabilities out of the remaining joint probabilities. Then the values
assigned to the rest of (out of the total of $64$) joint probabilities can
be found from (\ref{normalization-PD}) and (\ref{locality-1-PD}-\ref%
{locality-3-PD}) as $p_{2}=7/50,$ $p_{4}=1/100,$ $p_{7}=3/20,$ $%
p_{8}=21/100, $ $p_{14}=9/25,$ $p_{16}=1/10,$ $p_{24}=1/4,$ $p_{28}=9/50,$ $%
p_{31}=17/50,$ $p_{32}=7/25,$ $p_{36}=19/50,$ $p_{40}=31/50,$ $p_{47}=27/50,$
$p_{48}=23/50, $ $p_{54}=1/2,$ and $p_{56}=1/2$.

Now, for these values the left sides of the NE inequalities (\ref%
{(C,C,C)NE(A)}-\ref{(C,C,C)NE(C)}) are found as $106/1000,$ $96/1000,$ and $%
17/1000,$ respectively, showing that in this case the strategy profile $%
(C,C,C)$, consisting of the three players players playing the strategy of
Cooperation, is a NE. Note that this NE emerges because the resulting set of
joint probabilities is non-factorizable and that the constraints (\ref%
{Constraints on joint probs}) ensure that when this set becomes factorizable
this NE disappears and $(D,D,D)$ becomes the unique NE.

We notice that for a given set of numbers $\alpha /\beta ,$ $\theta /\beta ,$
$\delta /\theta ,$ $\omega /\beta $ (each being less than $1$) not every
non-factorizable set of joint probabilities can result in the strategy
triplet $(C,C,C)$ being a NE. That is, in this framework non-factorizability
is a necessary but not sufficient to make the strategy triplet $(C,C,C)$ to
be a NE. Also, in this paper we have not explored which other NE, apart from
the strategy triplet $(C,C,C)$, may emerge for a given non-factorizable set
of probabilities.

In the above approach considering three-player quantum game of PD we have
assumed that for a given set\footnote{%
It may, however, not be known whether such a set has been obtained
classically or quantum mechanically.} of joint probabilities, for which both
the \textit{normalization condition} and the \textit{causal communication
constraint} hold\footnote{%
as described by Cereceda \cite{Cereceda}.}, it is always possible in
principle to find a three-party two-choice EPR-Bohm setting, with the
relevant pure or mixed state(s) and appropriate directions of measurements
for the three observers, which can always generate the given set of joint
probabilities.

\subsubsection{Fate of the classical Nash equilibrium}

For factorizable joint probabilities the strategy profile $(D,D,D)$ is the
unique NE, represented by the strategy triplet $(x^{\star },y^{\star
},z^{\star })=(0,0,0)$. Here we show that this strategy profile remains a NE
even when joint probabilities may become non-factorizable. Using the payoff
relations (\ref{6-coin mixed-strategy payoffs}) and (\ref{6coin-B}) along
with the constraints (\ref{normalization-PD}, \ref{locality-1-PD}-\ref%
{locality-3-PD}), the NE conditions (\ref{6-coin NE}) for the strategy
profile $(D,D,D)$ for PD read

\begin{equation}
\begin{array}{l}
\Pi _{A}(0,0,0)-\Pi _{A}(x,0,0)=-x(\epsilon p_{36}+\omega p_{40}-\omega ),
\\ 
\Pi _{B}(0,0,0)-\Pi _{B}(0,y,0)=-y(\epsilon p_{47}+\omega p_{48}-\omega ),
\\ 
\Pi _{C}(0,0,0)-\Pi _{C}(0,0,z)=-z(\epsilon p_{54}+\omega p_{56}-\omega ).%
\end{array}%
\end{equation}%
Now, from (\ref{normalization-PD}) we have $p_{36}+p_{40}=1,$ $%
p_{47}+p_{48}=1,$ and $p_{54}+p_{56}=1.$ This makes the above three payoff
differences to be $xp_{36}(\omega -\epsilon ),$ $yp_{47}(\omega -\epsilon ),$
and $zp_{54}(\omega -\epsilon )$ for Alice, Bob, and Chris, respectively. As
for PD we have $\omega >\epsilon $ which makes these quantities non-negative
and the strategy profile $(D,D,D)$ remains a NE even when the joint
probabilities become non-factorizable. That is, a set of non-factorizable
joint probabilities can only add to the classical NE of $(D,D,D)$.

\section{Discussion}

This paper extends our probabilistic framework for playing two-player
quantum games to the multiplayer case. This framework presents an argument
for the construction of quantum games without referring to the tools of
quantum mechanics, thus making this area more accessible to workers outside
the quantum physics discipline.

Where does this framework stand in front of Enk and Pike's criticism \cite%
{EnkPike}, which refers to the standard setting for playing a quantum game
using Eisert et al.'s formalism? Their criticism considers a quantum game
equivalent to playing another classical game in which players have access to
extended set of classical strategies. This paper uses EPR-Bohm setting for
playing quantum games in which each player has two directions, along which a
measurement can be made by the referee, and his/her pure strategy, in a run,
consists of choosing one direction. A mixed strategy is then defined to be
the probability of playing one of his/her pure strategies. As the sets of
strategies remain exactly identical in both the classical and the quantum
forms of the game, it difficult to construct an Enk and Pike type argument 
\cite{EnkPike} for a quantum game played in the EPR-Bohm setting.

In the literature the wording `quantum games' has been used in many
different contexts, which involve games and quantum settings in one way or
the other. In view of this the present paper comes in line with the approach
towards quantum games that originated with Eisert et al.'s quantization of
Prisoner's Dilemma \cite{EWL}, where they proposed a general procedure to
quantize a given noncooperative two-player game. This approach is distinct,
and is to be contrasted, from other approaches that also use the wording
`quantum game', in that it is placed in the context of classical game theory
with the use of the Nash equilibrium concept. For example, in some
approaches both `the game' and its `winning condition(s)' are arbitrarily
defined, tailored or constructed, in order to show that only using a
quantum-mechanical implementation the winning condition(s) can be fulfilled.
In contrast, this paper begins by defining payoff relations, instead of the
`winning condition(s)', and then finds how game-theoretic outcome(s) of a
noncooperative game may change in relation to the quantum mechanical aspects
of a probabilistic physical system, which the players share in order to play
the game.

The approach followed in this paper establishes a relationship between the
`classicality' of the physical system (expressed by the joint probabilities
being factorizable) and a `classical game', in the sense that using a
classical system to play a game results in the classical game. Establishing
this relationship allows us in the following step to find how non-classical
(thus quantum) behavior of the physical system (expressed by the joint
probabilities being non-factorizable) may change the outcome(s) of the game.

This paper considers very unusual non-factorizable joint probabilities,
which may emerge from an entangled quantum state, by putting forward three-
and six-coin setups in order to play a three-player symmetric
non-cooperative game. In the six-coin setup, players choose coins in each
run, not necessarily unbiased, which are subsequently tossed and the outcome
of each toss is observed. This setup translates playing of the game in terms
of $64$ joint probabilities. In the following step, this translation of the
game allows us to consider the corresponding quantum game by bringing in the
same number of joint probabilities, which now may not be factorizable. We
then consider how these quantum mechanical probabilities may change the Nash
equilibria of the game under the constraint that factorizable joint
probabilities must lead to the classical solution of the game.

We achieve this by re-expressing players' payoffs in terms of $64$ joint
probabilities $p_{i}$, the players' strategies $x,$ $y,$ $z$\ and the
coefficients defining the game. We then use Nash inequalities to find the
equilibria. We find constraints on $p_{i}$ which ensure that for
factorizable $p_{i}$\ the game gives the classical outcome and thus it
becomes interpretable as a classical mixed-strategy game. This is carried
out by playing the game in the six-coin setup and using Nash inequalities to
obtain constraints on the coin probabilities $r,$\ $s,$\ $r^{\prime },$\ $%
s^{\prime },$ $r^{\prime \prime },$\ $s^{\prime \prime }$, which reproduce
the outcome of the classical mixed-strategy game. We use the relations (\ref%
{6Coin probs 1}-\ref{6Coin probs 8}), resulting from $p_{i}$ being
factorizable, to translate the constraints on $r,$\ $s,$\ $r^{\prime },$\ $%
s^{\prime },$ $r^{\prime \prime },$\ $s^{\prime \prime }$\ in terms of
constraints on $p_{i}$. We refer to the standard three-party EPR-Bohm setup
and allow $p_{i}$\ to be non-factorizable, while retaining the constraints
on $p_{i}$. We then ask if non-factorizability may lead to the emergence of
new solution(s) of the game. In case non-factorizability leads to new
solution(s), given as a triplet(s) $(x^{\ast },y^{\ast },z^{\ast }),$ there
is always is a set of $64$ non-factorizable joint probabilities that are
associated with it.

Our results show that non-factorizability leads to a new NE\ for
three-player PD that is pareto-optimal. As we recall from the Ref. \cite%
{IqbalCheon} that this does not turn out to be the case for the two-player
PD and the classical NE remains intact even when the players are given
access to non-factorizable probabilities. There exist, however, examples of
two-player games for which non-factorizability indeed leads to new NE. The
game of chicken \cite{Rasmusen} is one such example. It seems that from a
game-theoretic perspective the tri-partite quantum correlations are
different and stronger relative to the bi-partite quantum correlations.
Recall that non-factorizability, as it is defined above, is only a
necessary, but not sufficient, condition for the violation of Bell's
inequality. That is, a non-factorizable set of joint probabilities may not
violate Bell's inequality but a set of joint probabilities that violates
Bell's inequality must be non-factorizable. This might motivate one to ask
about the exact connection between non-factorizability, entanglement, and
the violation of Bell's inequality. However, non-factorizability being only
a necessary condition for the violation of Bell's inequality seems to go in
line with the known result that a separable mixed state can violate Bell's
inequality.

The probabilistic framework towards quantum games, developed in Ref.~\cite%
{IqbalCheon} and extended to multiplayer games in this paper, uses the
classical concept of probability, which is well known to be more restrictive
than the quantum notion \cite{Pitowsky}. Essentially, the classical concept
describes `probability'\ as being a number between zero and one and that for
a joint probability of two events this number is less or equal to the
numbers corresponding to each of the events. Though being more restrictive,
as Pitowski \cite{Pitowsky} describes it, this concept is "nevertheless
rooted in some very basic intuition." If quantum games can be expressed in
terms of this basic concept this can only be helpful to introduce this area
to researchers outside the quantum domain.

It is relevant here to point out that the probabilistic framework for
quantum games, proposed originally for two-player games in Ref.~\cite%
{IqbalCheon} and extended to three-player games in the present paper,
appears close to Einstein's statistical interpretation of quantum mechanics 
\cite{Ballentine,Jammer}. The key assertion of this interpretation describes
a quantum state (pure or otherwise), representing an ensemble of similarly
prepared systems, and need not provide a complete description of an
individual system. By using coin tosses in order to translate playing of a
classical game in terms of joint probabilities and subsequently introducing
unusual quantum mechanical non-factorizable joint probabilities, the
suggested framework uses the concept of an ensemble of similarly prepared
systems. Multiple coin tosses, which are central to the present framework,
are found helpful in understanding how quantum mechanical predictions in
quantum game theory do not pertain to a single measurement, but relate to an
ensemble of similar measurements.

\begin{acknowledgement}
One of us (AI) is supported at the University of Adelaide by the Australian
Research Council under the Discovery Projects scheme (Grant No. DP0771453).
Part of this work was carried out at the Kochi University of Technology
(KUT), Japan, where AI was supported by the Japan Society for Promotion of
Science (JSPS). At KUT this work was supported, in part, by the Grant-in-Aid
for scientific research provided by Japan Ministry of Education, Culture,
Sports, Science, and Technology under the contract numbers 18540384 and
18.06330.
\end{acknowledgement}

\end{document}